\documentclass[12pt, draftclsnofoot, onecolumn]{IEEEtran}
\usepackage{amssymb}
\usepackage{amsmath}
\usepackage{cite}
\usepackage{url}
\usepackage{xcolor}
\usepackage{cite,graphicx,amsmath,amssymb}
\usepackage{subfigure}
\usepackage{citesort}
\usepackage{fancyhdr}
\usepackage{mdwmath}
\usepackage{mdwtab}
\usepackage{caption}
\usepackage{amsthm}
\usepackage{setspace}
\usepackage{algorithm}
\usepackage{algorithmic}
\usepackage{makecell}
\usepackage{diagbox}
\newcommand{\bm}[1]{\mbox{\boldmath{$#1$}}}

\newtheorem{theorem}{Theorem}

\newtheorem{lemma}{Lemma}

\newtheorem{corollary}{Corollary}

\allowdisplaybreaks
\makeatletter
\def\ScaleIfNeeded{%
\ifdim\Gin@nat@width>\linewidth \linewidth \else \Gin@nat@width
\fi } \makeatother

\begin{document}
%\title{Multiple Access for Deployment-aware Intelligent Reflecting Surface Assisted Networks}
%\title{Multiple Access for Intelligent Reflecting Surface Assisted Networks: A Deployment Optimization Perspective}
\title{Intelligent Reflecting Surface Enhanced Indoor Robot Path Planning: A Radio Map based Approach}
%
%Trajectory Design for Mission Completion Time Minimization in Cellular-enabled UAV Uplink NOMA Communication
%\author{
%\IEEEauthorblockN{ Yuanwei~Liu\IEEEauthorrefmark{1}, Zhijin~Qin\IEEEauthorrefmark{1}, Maged Elkashlan\IEEEauthorrefmark{1}, and  Yue~Gao\IEEEauthorrefmark{1}\\} \IEEEauthorblockA{
%\IEEEauthorrefmark{1} Queen Mary University of London, London, UK\\
%%\IEEEauthorrefmark{2} Lancaster University, Lancaster, UK\\
% } }

\author{

Xidong~Mu,~\IEEEmembership{Student Member,~IEEE,}
        Yuanwei~Liu,~\IEEEmembership{Senior Member,~IEEE,}
       Li~Guo,~\IEEEmembership{Member,~IEEE,}
       Jiaru~Lin,~\IEEEmembership{Member,~IEEE,}
       and Robert~Schober,~\IEEEmembership{Fellow,~IEEE}

\thanks{Part of this work has been submitted to the IEEE International Conference on Communications (ICC), Montreal, Canada, June 14-18, 2021.~\cite{Mu2021icc}}
%\thanks{Part of this work has been submitted to the IEEE International Conference on Communications Workshop on NOMA for 5G and Beyond, Dublin, Ireland, June 7-11, 2020.~\cite{Mu2020WS}}
\thanks{X. Mu, L. Guo, and J. Lin are with the School of Artificial Intelligence and the Key Laboratory of Universal Wireless Communications, Ministry of Education, Beijing University of Posts and Telecommunications, Beijing, China. (email:\{muxidong, guoli, jrlin\}@bupt.edu.cn).}
\thanks{Y. Liu is with the School of Electronic Engineering and Computer Science, Queen Mary University of London, London, UK. (email:yuanwei.liu@qmul.ac.uk).}
\thanks{R. Schober is with the Institute for Digital Communications, Friedrich-Alexander-University Erlangen-N{\"u}rnberg (FAU), Germany (e-mail: robert.schober@fau.de).}
}

\maketitle
\vspace{-1.5cm}
\begin{abstract}
\vspace{-0.3cm}
\textcolor{black}{Integrating robots into cellular networks creating connected robotic users has emerged as a promising technology for future smart cities and smart factories due to their low cost and high maneuverability. However, the requirement of establishing stable and high-quality communication links to the robotic users greatly restricts their applicability, especially in indoor environments where obstacles may block the wireless link. To tackle this challenge, in this paper,} an indoor robot navigation system is investigated, where an intelligent reflecting surface (IRS) is employed to enhance the connectivity between the access point (AP) and robotic users. Both single-user and multiple-user scenarios are considered. In the single-user scenario, one mobile robotic user (MRU) communicates with the AP. In the multiple-user scenario, the AP serves one MRU and one static robotic user (SRU) employing either non-orthogonal multiple access (NOMA) or orthogonal multiple access (OMA) transmission. The considered system is optimized for minimization of the travelling time/distance of the MRU from a given starting point to a predefined final location, while satisfying constraints on the communication quality of the robotic users. To this end, a \emph{radio map} based approach is proposed to exploit location-dependent channel propagation knowledge. For the single-user scenario, a \emph{channel power gain map} is constructed, which characterizes the spatial distribution of the maximum expected effective channel power gain of the MRU for the optimal IRS phase shifts. Based on the obtained channel power gain map, the communication-aware robot path planing problem is solved by exploiting graph theory. For the multiple-user scenario, a \emph{communication rate map} is constructed, which characterizes the spatial distribution of the maximum expected rate of the MRU for the optimal power allocation at the AP and the optimal IRS phase shifts subject to a minimum rate requirement for the SRU. The joint optimization problem is efficiently solved by invoking bisection search and successive convex approximation methods. Then, a graph theory based solution for the robot path planning problem is derived by exploiting the obtained communication rate map. Our numerical results show that: 1) the required travelling distance of the MRU can be significantly reduced by deploying an IRS; 2) NOMA yields a higher communication rate for the MRU than OMA; 3) the IRS performance gain is significantly more pronounced for NOMA than for OMA.
\end{abstract}
%\vspace{-0.6cm}
%\begin{keywords}
%\vspace{-0.3cm}
%Intelligent reflecting surfaces, non-orthogonal multiple access, radio maps, robot path planning.
%\end{keywords}
%\vspace{-0.4cm}
\section{Introduction}
\vspace{-0.0cm}
In the past few decades, robot technology has developed rapidly and has had a significant impact on human life \cite{Handbook_robot}. Specifically, robots can help humans perform repetitive or dangerous tasks, thus liberating human resources and reducing health risks. There is a wide range of robot applications, including cargo/packet delivery, search and rescue, public safety surveillance, environmental monitoring, and automatic industrial production \cite{robot_cloud,Robots_EM}. In terms of their modes of operation, current robots can be loosely classified into two categories, namely, automated robots and connected robots \cite{connected_robot}. Based on the equipped sensors and computational resources, automated robots are able to make decisions on their own during a mission. However, they are exceedingly complex due to the large memory, large computational resources, and large number of artificial intelligence based algorithms needed for carrying out sophisticated tasks. In contrast, connected robots accomplish missions relying on information exchange with operators \cite{connected_robot}. For instance, a connected robot sends the sensed data (e.g., pictures or videos) to its operator in a real-time manner, and the operator provides further instructions to the connected robot based on the data. Therefore, connected robots are more cost-efficient and less computation-constrained. With the rapid development of fifth-generation (5G) and beyond (B5G) cellular networks, one promising solution is to integrate connected robots into cellular networks as robotic users to be served by base stations (BSs) or access points (APs). Given the ultra-high speed, low latency, and high reliability of 5G/B5G networks, connected robots are expected to become a key technology in the future.\\
\indent Despite the appealing advantages of connected robots, one crucial limitation is that the communication link may be severely blocked by buildings, trees or other tall objects. The resulting signal \emph{dead zones} can significantly restrict the area of operation and reduce the efficiency of connected robots. Fortunately, with the recent advances in meta-materials, intelligent reflecting surfaces (IRSs) \cite{WuTowards}, also known as reconfigurable intelligent surfaces (RISs) \cite{RIS_survey,Renzo_IRS} or large intelligent surfaces (LISs) \cite{LiangLISA}, have been proposed as an effective solution for overcoming signal blockage and enhancing the communication quality. An IRS is a thin man-made surface consisting of a large number of low-cost and passive reflecting elements, each of which can reflect and impact the propagation of an incident electromagnetic wave \cite{WuTowards}. As a result, IRSs can create a \emph{programmable wireless environment}. If the signal transmission via the direct link is blocked, an IRS can be deployed to provide an additional reflected link, hence improving the received signal strength. As the IRS does not require radio frequency (RF) chains and only reflects the incident signal in a nearly passive manner, it is more cost- and energy-efficient than conventional relaying technologies such as amplify-and-forward (AF) and decode-and-forward (DF) relaying \cite{Huang_Mag}. Furthermore, IRSs can be easily deployed on different structures, such as building facades and roadside billboards in outdoor environments, as well as walls and ceilings in indoor environments.
\vspace{-1.4cm}
\subsection{Prior Works}
\vspace{-1cm}
\textcolor{black}{\subsubsection{Robot Path Planning} For the application of robots, path planning is essential for robots to carry out tasks reliably and safely. Hence, the robot path planning problem has been studied extensively in the past few decades. To ensure that robots will not collide with obstacles in their workspace, prior studies on robot path planning have proposed different algorithms for different application scenarios. By discretizing the continuous space into a finite grid, efficient robot path planning algorithms, including the Dijkstra, A*, and D* algorithms, were developed to find the shortest path between two locations in static and dynamic environments~\cite{Graph_robot}. To cope with the more stringent challenges introduced by uncertain environments, the authors of \cite{PSO} proposed a particle swarm optimization (PSO) based robot path planning algorithm. The authors of \cite{ABC} investigated the multi-robot path planning problem and proposed an artificial bee colony optimization algorithm to minimize the sum path length of all robots. Exploiting machine learning, the authors of \cite{DRL_robot} invoked deep reinforcement learning for collision avoidance.}
\subsubsection{IRS-assisted Communication System Design} IRSs have received extensive attention from both academia and industry recently. By exploiting the new degrees of freedom introduced by passive beamforming, the performance gain facilitated by IRSs in wireless communication systems has been extensively investigated. For instance, the authors of \cite{Wu2019IRS} proposed an alternating optimization based algorithm for the design of the active beamforming at the BS and the passive beamforming at the IRS with the objective of minimizing the transmit power. The authors of \cite{Huang_EE} investigated energy-efficiency maximization in an IRS-assisted multiple-user multiple-input single-output (MISO) system. In \cite{Yu_secure}, the authors studied the physical layer security in IRS-aided communication systems, where the system sum secrecy rate was maximized. In \cite{Huang_indoor}, the authors considered an indoor IRS communication scenario, where the IRS phase shifts were configured by the proposed deep learning method to maximize the user's received signal strength. The authors of \cite{Pan_IRS} focused on a multi-cell multiple-input multiple-output (MIMO) multiple-user communication system, where an IRS was deployed at the cell boundary to improve the performance of the cell-edge users. Furthermore, the authors of \cite{Huang_IRS_DL} invoked deep reinforcement learning techniques to tackle the joint active and passive beamforming problem. The proposed algorithm was shown to be capable of learning from the environment. In \cite{IRS_UAV}, the authors investigated IRS-assisted unmanned aerial vehicle (UAV) communication, where the UAV trajectory and the IRS phase shifts were jointly optimized to maximize the average achievable rate of a ground user. To further improve spectrum efficiency, non-orthogonal multiple access (NOMA) was considered for IRS-assisted communication systems. The authors of \cite{Hou_IRS} analyzed various system performance metrics in an IRS-aided NOMA system, and provided useful design insights. With the aim of maximizing the system sum rate, joint active and passive beamforming optimization was investigated in \cite{Mu_IRS} for IRS-assisted MISO NOMA communication systems.
%A similar UAV trajectory design problem was also considered in \cite{RIS_UAV2} for multiple IRSs.
\vspace{-0.6cm}
\subsection{Motivations and Challenges}
\begin{figure}[t!]
    \begin{center}
        \includegraphics[width=2.4in]{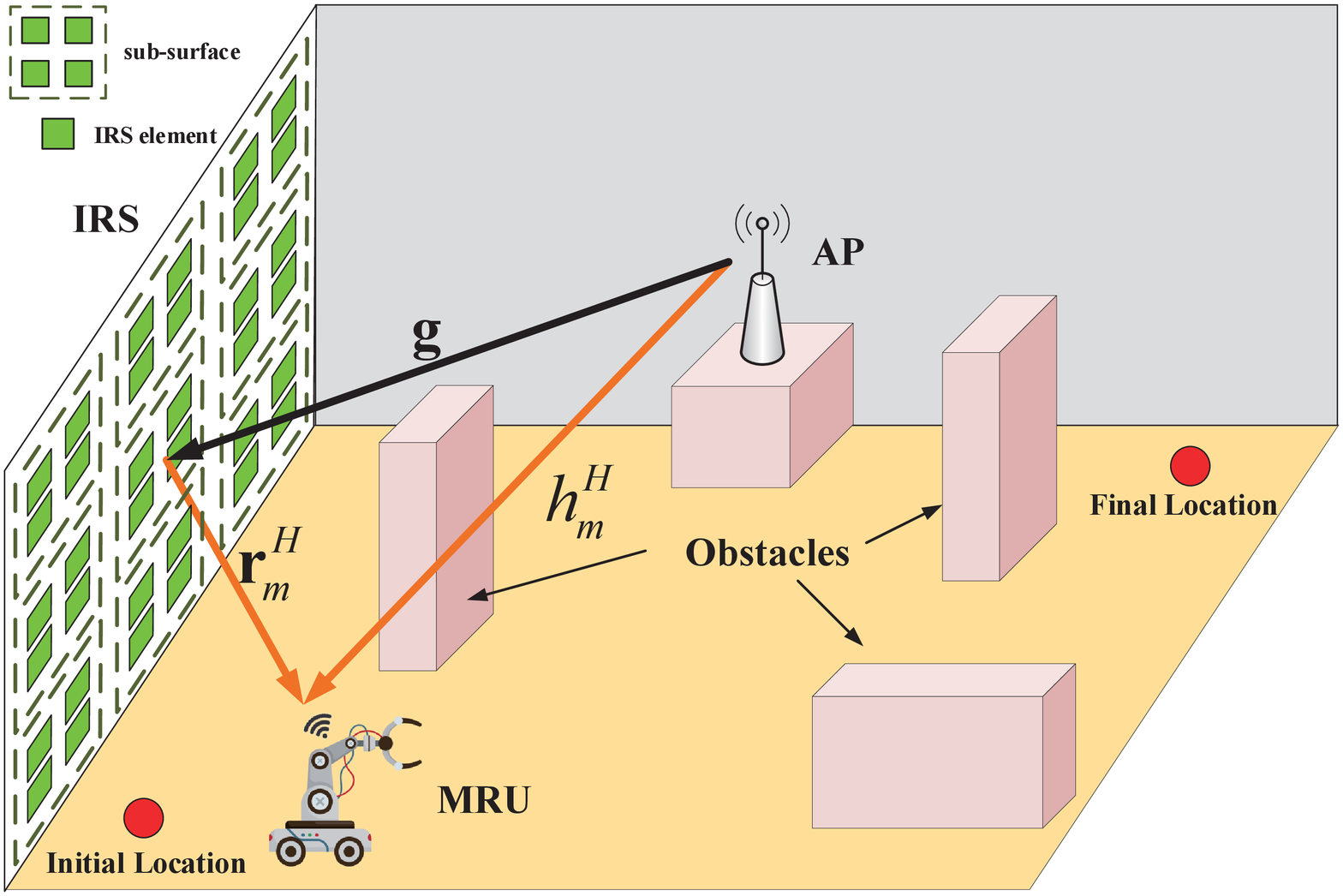}
        \caption{Illustration of the IRS-enhanced indoor robot navigation system for a single-user scenario.}
        \label{System model}
    \end{center}
\end{figure}

\vspace{-0.1cm}
On the road to facilitating smart cities and factories, connected robots have been regarded as an appealing technology. By offloading tasks to remote operators (e.g., BSs or APs), the cost and energy consumption of robots can be significantly reduced. However, as mentioned earlier, signal blockage is a major bottleneck for connected robots. Motivated by this issue, we propose to deploy an IRS to assist the communication with a connected robot. In particular, an IRS-enhanced indoor connected robot navigation system is considered, where one mobile robotic user (MRU) is served by an AP with the aid of an IRS, see Fig. \ref{System model}. The MRU is dispatched to travel from a predefined initial location to a final location to carry out a specific mission. \textcolor{black}{During the travel, the MRU's instantaneous communication quality should not fall below a certain threshold to prevent the loss of control.} Although the performance gain introduced by IRSs has been studied for various wireless communication system architectures, to the best of the authors' knowledge, this is the first work to investigate the communication-aware indoor robot path planning problem in IRS-enhanced environments. The main related challenges are as follows:
\begin{itemize}
  \item \textcolor{black}{As the wireless link may be blocked by obstacles, as illustrated in Fig. 1, the communication quality of the MRU may change abruptly as it travels. This makes the considered communication-aware indoor robot path planning problem more challenging than the conventional robot path planning problem which is only concerned with collision avoidance.}
  \item In addition, the communication quality of the MRU does not only depend on its location but also on the IRS phase shifts, which causes the path planning and the IRS reflection matrix design to be highly coupled.
\end{itemize}
To overcome the aforementioned challenges, we develop a new radio-map based approach for the robot path planning problem. In general, a radio map contains information on the spectral activities and the propagation conditions in the space, frequency, and time domains \cite{radiomap}. This information can be exploited to improve the performance of wireless networks and facilitates new wireless applications. Inspired by this, we construct two types of radio maps, namely, a \emph{channel power gain map} and a \emph{communication rate map} for single-user and multiple-user scenarios, respectively, where we exploit knowledge about the channel propagation conditions. In the single-user case, one MRU is assumed to be served by the AP in a dedicated resource block with the aid of an IRS. As the communication performance is fully determined by the channel quality, a channel power gain map is constructed to characterize the maximum expected channel power gain of the MRU in the region of interest. In the multiple-user case, we consider one MRU and one static robotic user (SRU) which are simultaneously served by the AP. The communication performance of the robotic users also depends on the resource allocation at the AP. Hence, we construct a communication rate map, which characterizes the spatial distribution of the maximum expected rate of the MRU. Here, we jointly consider the MRU's location, the power allocation at the AP, and the phase shifts at the IRS. Equipped with the two radio maps, the communication-aware robot path planning problem is efficiently solved by utilizing graph theory.
\vspace{-0.6cm}
\subsection{Contributions}
\vspace{-0.3cm}
\indent The main contributions of this paper can be summarized as follows:
\begin{itemize}
  \item We propose an IRS-enhanced indoor robot navigation system, in which the IRS is deployed to enhance the signal transmission from the AP to an MRU. We formulate a communication-aware path planning problem to minimize the time/distance needed by the MRU for travelling from an initial to a final location. In particular, we jointly optimize the robot path and the IRS reflection matrix. \emph{Radio map} \cite{radiomap} based approaches are proposed for both single-user and multiple-user systems.
  \item For the single-user scenario, we first construct the \emph{channel power gain map} by optimizing the IRS phase shifts. The obtained channel power gain map characterizes the spatial distribution of the maximum expected effective channel power gain of the MRU aided by the IRS. Leveraging this map, the robot path planning problem is efficiently solved using graph theory.
  \item For the multiple-user scenario, we consider both NOMA and orthogonal multiple access (OMA) transmission schemes for simultaneously serving one MRU and one SRU. The \emph{communication rate map} of the MRU is constructed by jointly optimizing the power allocation at the AP and the reflection matrix at the IRS, subject to the rate constraint of the SRU. Specifically, we solve the resulting joint optimization problem by invoking the bisection search and successive convex approximation (SCA) methods. Based on the rate map, a graph theory based solution for the path of the MRU is obtained.
  \item We show that the proposed IRS-enhanced system can significantly reduce the travelling distance of the MRU while achieving a higher communication rate compared to conventional systems without IRS. We also show that NOMA outperforms OMA, especially for the IRS-enhanced system.
\end{itemize}
\vspace{-0.6cm}
\subsection{Organization and Notation}
\vspace{-0.2cm}
The rest of this paper is organized as follows: In Section II, IRS-enhanced indoor robot path planning for a single-user system is investigated, and a channel power gain map is constructed for solving the formulated problem. In Section III, the robot path planning problem is extended to multiple-user systems for both NOMA and OMA transmission, and a communication rate map is constructed. Numerical examples are presented in Section IV to verify the effectiveness of the proposed designs compared to benchmark schemes. Finally, Section VI concludes the paper. \\
\indent \emph{Notations:} Scalars, vectors, and matrices are denoted by lower-case, bold-face lower-case, and bold-face upper-case letters, respectively. ${\mathbb{C}^{N \times 1}}$ denotes the space of $N \times 1$ complex-valued vectors. The transpose and conjugate transpose of vector ${\mathbf{a}}$ are denoted by ${{\mathbf{a}}^T}$ and ${{\mathbf{a}}^H}$, respectively. ${\left\| {\mathbf{a}} \right\|_1}$ and $\left\| {\mathbf{a}} \right\|$ denote the 1-norm and the Euclidean norm of vector ${\mathbf{a}}$, respectively. ${\rm {diag}}\left( \mathbf{a} \right)$ denotes a diagonal matrix with the elements of vector ${\mathbf{a}}$ on the main diagonal. ${{\mathbf{1}}_{m \times n}}$ denotes an all-one matrix of size ${m \times n}$. ${\mathbb{H}^{N}}$ denotes the set of all $N$-dimensional complex Hermitian matrices. ${\rm {rank}}\left( \mathbf{A} \right)$ and ${\rm {Tr}}\left( \mathbf{A} \right)$ denote the rank and the trace of matrix $\mathbf{A}$, respectively. ${{\mathbf{A}}} \succeq 0$ indicates that $\mathbf{A}$ is a positive semidefinite matrix. $\otimes $ denotes the Kronecker product. ${\left[ \cdot  \right]_n}$ denotes the $n$th element of a vector.
\vspace{-0.3cm}
\section{Radio Map based Robot Path Planning for Single-user System}
\vspace{-0.1cm}
\subsection{System Model}
\vspace{-0.1cm}
In this section, we consider an IRS-enhanced indoor robot navigation system, which consists of one single-antenna AP, one single-antenna MRU, and one IRS with $M$ passive reflecting elements, see Fig. \ref{System model}. The IRS is deployed on one of the indoor walls for assisting the transmission from the AP to the robotic user. Adopting a three-dimensional (3D) Cartesian coordinate system, the locations of the antenna of the AP and \textcolor{black}{the center of the IRS} are denoted by ${\mathbf{b}}=\left( {x_b,y_b,{H_b}} \right)$ and ${\mathbf{f}}=\left( {x_f,y_f,{H_f}} \right)$, respectively. The MRU is dispatched to travel from an initial location ${{\mathbf{q}}_I} = \left( {{x_I},{y_I},{H_0}} \right)$ to a final location ${{\mathbf{q}}_F} = \left( {{x_F},{y_F},{H_0}} \right)$, where ${{H_0}}$ denotes the height of the antenna of the MRU. Let ${\mathbf{q}}\left( t \right) = \left( {x\left( t \right),y\left( t \right),{H_0}} \right),t \in \left[ {0,T} \right]$, denote the time-varying path of the MRU, where $T$ denotes the required travelling time\footnote{The considered setup is representative for many practical connected robot applications, such as transportation of material in smart factories or delivery of medicine in hospitals.}. For practical implementation, the IRS is equipped with a smart controller, realized, e.g., with a field-programmable gate array (FPGA), which allows the AP to configure the IRS phase shifts in a real time manner. As the AP-IRS-user link suffers from severe path loss, a large number of reflecting elements are required for this link to achieve a comparable path loss as the unobstructed  direct AP-user link \cite{tile}. However, a large number of reflecting elements also cause a prohibitively high overhead/complexity for channel acquisition and phase shift design/reconfiguration. To overcome this limitation, an effective method is to group adjacent reflecting elements, which are expected to experience high channel correlation, together to a sub-surface, as was done in \cite{Yang_IRS}. All elements belonging to the same sub-surface are assumed to share the same reflection coefficient. In this paper, the $M$ passive reflecting elements of the IRS are divided into $N$ sub-surfaces, where each sub-surface consists of $\overline N  = {M \mathord{\left/
 {\vphantom {M N}} \right.
 \kern-\nulldelimiterspace} N}$ reflecting elements. An example where $\overline N =4$ elements are grouped into a sub-surface is illustrated in Fig. \ref{System model}. The instantaneous IRS reflection matrix is denoted by ${\mathbf{\Theta }}\left( t \right) = {\rm{diag}}\left( {{\bm{\theta }}\left( t \right) \otimes {{\mathbf{1}}_{\overline N \times 1 }}} \right)\in {\mathbb{C}^{M \times M}}$, where ${\bm{\theta }}\left( t \right) = {\left[ {{\beta _1}\left( t \right){e^{j{\theta _1}\left( t \right)}},{\beta _2}\left( t \right){e^{j{\theta _2}\left( t \right)}}, \ldots ,{\beta _N}\left( t \right){e^{j{\theta _N}\left( t \right)}}} \right]^T}$, and ${{\theta _n}\left( t \right)}$ and ${{\beta _n}\left( t \right)}$ denote the instantaneous phase shift and attenuation coefficient of the $n$th sub-surface of the IRS, respectively. In this paper, we assume ${\theta _n}\left( t \right) \in \left[ {0,2\pi } \right)$ and ${\beta _n}\left( t \right) = 1,\forall n \in {\mathcal{N}},t \in {{\mathcal{T}}}$, where ${\mathcal{N}} = \left\{ {1, \ldots ,N} \right\}$ and ${{\mathcal{T}}} = \left[ {0,T} \right]$.\\
\indent We focus our attention on the downlink transmission from the AP to the MRU. The channel between the AP and the IRS is denoted by ${\mathbf{g}} \in {\mathbb{C}^{M \times 1}}$, and follows the Rician channel model. Hence, ${\mathbf{g}}$ can be expressed as
\begin{align}\label{AP-IRS}
{\mathbf{g}} = \frac{{\sqrt {{{\mathcal{L}}_{AI}}} }}{{\sqrt {{K_{AI}} + 1} }}\left( {\sqrt {{K_{AI}}} \overline {\mathbf{g}}  + \widehat {\mathbf{g}}} \right),
\end{align}
where ${{{\mathcal{L}}_{AI}}}$ is the distance-dependent path loss of the AP-IRS channel, $\overline {\mathbf{g}}$ denotes the deterministic line-of-sight (LoS) component, $\widehat {\mathbf{g}}$ denotes the random non-LoS (NLoS) component, which follows the Rayleigh distribution, and $K_{AI}$ is the Rician factor.\\
\indent Furthermore, let ${h_m}\left( {{\mathbf{q}}\left( t \right)} \right)\in {\mathbb{C}^{1 \times 1}}$ and ${{\mathbf{r}}_m}\left( {{\mathbf{q}}\left( t \right)} \right)\in {\mathbb{C}^{M \times 1}}$ denote the AP-MRU and IRS-MRU channels for MRU location ${\mathbf{q}}\left( t \right)$. We have
\vspace{-0.3cm}
\begin{align}\label{AP-robotic}
{h_m}\left( {{\mathbf{q}}\left( t \right)} \right) = \frac{{\sqrt {{{\mathcal{L}}_{AM}}\left( {{\mathbf{q}}\left( t \right)} \right)} }}{{\sqrt {{K_{AM}}\left( {{\mathbf{q}}\left( t \right)} \right) + 1} }}\left( {\sqrt {{K_{AM}}\left( {{\mathbf{q}}\left( t \right)} \right)} {{\overline h}_m}\left( {{\mathbf{q}}\left( t \right)} \right) + {{\widehat h}_m}} \right),
\end{align}
\vspace{-1cm}
\begin{align}\label{IRS-robotic}
{{\mathbf{r}}_m}\left( {{\mathbf{q}}\left( t \right)} \right) = \frac{{\sqrt {{{\mathcal{L}}_{IM}}\left( {{\mathbf{q}}\left( t \right)} \right)} }}{{\sqrt {{K_{IM}}\left( {{\mathbf{q}}\left( t \right)} \right) + 1} }}\left( {\sqrt {{K_{IM}}\left( {{\mathbf{q}}\left( t \right)} \right)} {{\overline {\mathbf{r}} }_m}\left( {{\mathbf{q}}\left( t \right)} \right) + {{\widehat {\mathbf{r}}}_m}} \right),
\end{align}
\vspace{-1cm}

\noindent where ${{{\mathcal{L}}_{AM}}\left( {{\mathbf{q}}\left( t \right)} \right)}$ and ${{{\mathcal{L}}_{IM}}\left( {{\mathbf{q}}\left( t \right)} \right)}$ denote the corresponding path losses. ${{{\overline h}_m}\left( {{\mathbf{q}}\left( t \right)} \right)}$ and ${{{\overline {\mathbf{r}} }_m}\left( {{\mathbf{q}}\left( t \right)} \right)}$ are the location-dependent LoS components. ${{\widehat h}_m}$ and ${\widehat {\mathbf{r}}_m}$ denote the random Rayleigh distributed NLoS components. ${{K_{AM}}\left( {{\mathbf{q}}\left( t \right)} \right)}$ and ${{K_{IM}}\left( {{\mathbf{q}}\left( t \right)} \right)}$ denote the location-dependent Rician factors. For instance, if the wireless link between the MRU at location ${{\mathbf{q}}\left( t \right)}$ and the AP/IRS is blocked by obstacles, the corresponding channel is classified as NLoS and we have ${K_{AM/IM}}\left( {{\mathbf{q}}\left( t \right)} \right) = 0$\footnote{\textcolor{black}{Note that, in practice, even if the direct LoS link is blocked, some of its energy may be reflected by the ceiling, floor, as well as walls and arrive at the receiver. However, we ignore the LoS link energy reflected by objects other than the IRS, which is a worst-case assumption for actual communication system design. Nevertheless, the results of this paper can be readily extended to the scenario with non-negligible reflected LoS links by adopting the corresponding channel models for radio map design.}}. Otherwise, it is classified as an LoS dominated channel and ${K_{AM/IM}}\left( {{\mathbf{q}}\left( t \right)} \right) = {\kappa _{AM/IM}}$, where ${\kappa _{AM/IM}}$ is a constant. \\
\indent Due to the high path loss, similar to \cite{Wu2019IRS}, signals that are reflected by the IRS two or more times are ignored. Therefore, the IRS-aided effective channel between the AP and the MRU can be expressed as
\vspace{-0.3cm}
\begin{align}\label{effective static}
{c_m}\left( t \right) = h_m^H\left( {{\mathbf{q}}\left( t \right)} \right) + {\mathbf{r}}_m^H\left( {{\mathbf{q}}\left( t \right)} \right){\mathbf{\Theta}} \left( t \right){\mathbf{g}}.
\end{align}
\vspace{-1cm}

\noindent We note that ${c_m}\left( t \right)$ is a random variable since it depends on random variables $\left\{ {\widehat {\mathbf{g}},{{\widehat {\mathbf{r}}}_m},{{\widehat h}_m}} \right\}$. In this paper, we are interested in the expected/average effective channel power gain, defined as ${\mathbb{E}}\left[ {{{\left| {{c_m}\left( t \right)} \right|}^2}} \right]$. A closed-form expression for ${\mathbb{E}}\left[ {{{\left| {{c_m}\left( t \right)} \right|}^2}} \right]$ is provided in the following lemma.
\begin{lemma}\label{expected effective channel power gain}
\emph{The expected effective channel power gain of the MRU is given by}
\begin{align}\label{expected robotic channel gain0}
\begin{gathered}
  {\mathbb{E}}\left[ {{{\left| {{c_m}\left( t \right)} \right|}^2}} \right] \triangleq {\lambda _m}\left( t \right) =\! {\left| {\widetilde h_m^H\!\left( {{\mathbf{q}}\left( t \right)} \right)\! +\! {{\widetilde {\mathbf{r}}}_m^H}\!\left( {{\mathbf{q}}\left( t \right)} \right)\!{\mathbf{\Theta}} \left( t \right)\!\widetilde {\mathbf{g}}} \right|^2}\!\hfill \\
    +\! \frac{{{{\mathcal{L}}_{AM}}\!\left( {{\mathbf{q}}\left( t \right)} \right)}}{{{K_{AM}}\!\left( {{\mathbf{q}}\left( t \right)} \right) \!+\! 1}}\!+\! \frac{{{{{\mathcal{L}}_{AI}}}{{{\mathcal{L}}_{IM}}\!\left( {{\mathbf{q}}\left( t \right)} \right)}\left( {{K_{IM}}\!\left( {{\mathbf{q}}\left( t \right)} \right)\! +\! {K_{AI}}\! +\! 1} \right)M}}{\left( {{K_{AI}}\! +\! 1} \right){\left( {{K_{IM}}\!\left( {{\mathbf{q}}\left( t \right)} \right)\! +\! 1} \right)}}, \hfill \\
\end{gathered}
\end{align}
\emph{where} $\widetilde h_m^H\left( {{\mathbf{q}}\left( t \right)} \right) = \sqrt {\frac{{{{{\mathcal{L}}_{AM}}\left( {{\mathbf{q}}\left( t \right)} \right)}{K_{AM}}\left( {{\mathbf{q}}\left( t \right)} \right)}}{{{K_{AM}}\left( {{\mathbf{q}}\left( t \right)} \right) + 1}}} \overline h_m^H{\left( {{\mathbf{q}}\left( t \right)} \right)},$ ${\widetilde {\mathbf{r}}_m^H}\left( {{\mathbf{q}}\left( t \right)} \right) = \sqrt {\frac{{{{{\mathcal{L}}_{IM}}\left( {{\mathbf{q}}\left( t \right)} \right)}{K_{IM}}\left( {{\mathbf{q}}\left( t \right)} \right)}}{{{K_{IM}}\left( {{\mathbf{q}}\left( t \right)} \right) + 1}}} {\overline {\mathbf{r}} _m^H}\left( {{\mathbf{q}}\left( t \right)} \right),$ $\widetilde {\mathbf{g}} = \sqrt {\frac{{{{\mathcal{L}}}_{AI}}{{K_{AI}}}}{{{K_{AI}} + 1}}} \overline {\mathbf{g}}$.
\begin{proof}
See Appendix~A.
\end{proof}
\end{lemma}
\indent \textcolor{black}{Based on \textbf{Lemma \ref{expected effective channel power gain}}, the calculation of ${\lambda _m}\left( t \right)$ requires the statistical channel state information (S-CSI) of each link, which can be obtained as follows: First, the instantaneous CSI (I-CSI) of each link is estimated several times via one of the recently proposed CSI estimation methods for IRS~\cite{Yang_IRS,Zheng_OFDM_WCL}. Then, exploiting the estimated I-CSI, the S-CSI is obtained by employing standard mean and covariance matrix estimation techniques~\cite{Mestre,Werner}.} For ease of exposition, let ${\mathbf{w}}_m^H\left( {{\mathbf{q}}\left( t \right)} \right) = \widetilde {\mathbf{r}}_m^H\left( {{\mathbf{q}}\left( t \right)} \right){\rm{diag}}\left( {\widetilde {{\mathbf{g}}}} \right) \in {\mathbb{C}^{1 \times M}}$ denote the cascaded LoS channel of the AP-IRS-MRU link before the reconfiguration of the IRS. Then, the corresponding combined composite channel associated with the $n$th sub-surface is given by ${\left[ {{{\widetilde {\mathbf{w}}}_m^H}\left( {{\mathbf{q}}\left( t \right)} \right)}\in {\mathbb{C}^{1 \times N}} \right]_n} = \sum\nolimits_{\overline n = 1}^{\overline N} {{{\left[ {{{\mathbf{w}}_m^H}\left( {{\mathbf{q}}\left( t \right)} \right)} \right]}_{\overline n + \left( {n - 1} \right)\overline N}}} ,\forall n \in {{\mathcal{N}}}$~\cite{Yang_IRS}. Therefore, ${\lambda _m}\left( t \right)$ can be rewritten as
\vspace{-0.3cm}
\begin{align}\label{expected robotic channel gain}
\begin{gathered}
  {\lambda _m}\left( t \right) = {\left| {\widetilde h_m^H\left( {{\mathbf{q}}\left( t \right)} \right) + {\mathbf{w}}_m^H\left( {{\mathbf{q}}\left( t \right)} \right)\left( {{\bm{\theta }}\left( t \right) \otimes {{\mathbf{1}}_{\overline N  \times 1}}} \right)} \right|^2}+{\tau _m}\left( {{\mathbf{q}}\left( t \right)} \right) \hfill \\
   \;\;\;\;\;\;\;\;\;\;= {\left| {\widetilde h_m^H\left( {{\mathbf{q}}\left( t \right)} \right) + \widetilde {\mathbf{w}}_m^H\left( {{\mathbf{q}}\left( t \right)} \right){\bm{\theta }}\left( t \right)} \right|^2}+{\tau _m}\left( {{\mathbf{q}}\left( t \right)} \right), \hfill \\
\end{gathered}
\end{align}
\vspace{-0.5cm}

\noindent where ${\tau _m}\left( {{\mathbf{q}}\left( t \right)} \right) = \frac{{{{{\mathcal{L}}}_{AM}}\left( {{\mathbf{q}}\left( t \right)} \right)}}{{{K_{AM}}\left( {{\mathbf{q}}\left( t \right)} \right) + 1}} + \frac{{{{{\mathcal{L}}}_{AI}}{{{\mathcal{L}}}_{IM}}\left( {{\mathbf{q}}\left( t \right)} \right)\left( {{K_{IM}}\left( {{\mathbf{q}}\left( t \right)} \right) + {K_{AI}} + 1} \right)M}}{{\left( {{K_{AI}} + 1} \right)\left( {{K_{IM}}\left( {{\mathbf{q}}\left( t \right)} \right) + 1} \right)}}$.
\vspace{-0.5cm}
\subsection{Problem Formulation}
\vspace{-0.1cm}
\indent We aim to minimize the required travelling time $T$ of the MRU from ${{\mathbf{q}}_I}$ to ${{\mathbf{q}}_F}$ by jointly optimizing the path of the MRU, ${{Q}} = \left\{ {{\mathbf{q}}\left( t \right),0 \le t \le T} \right\}$, and the reflection matrix of the IRS $A = \left\{ {{\mathbf{\Theta}} \left( t \right),0 \le t \le T} \right\}$, subject to a constraint on the expected effective channel power gain. Hence, the communication-aware robot path planning problem can be formulated as
\vspace{-0.5cm}
\begin{subequations}\label{P_single}
\begin{align}
&\mathop {\min }\limits_{Q,A,T} \;T\\
\label{single user rate requirment2}{\rm{s.t.}}\;\;&{\lambda _m}\left( t \right) \ge \overline \gamma  ,\forall t \in {{\mathcal{T}}},\\
\label{IRS phase}&{\theta _n}\left( t \right) \in \left[ {0,2\pi } \right),\forall n \in {\mathcal{N}},t \in {{\mathcal{T}}},\\
\label{Initial Location Constraint}&{{\mathbf{q}}}\left( 0 \right) = {\mathbf{q}}_I,{{\mathbf{q}}}\left( T \right) = {\mathbf{q}}_F,\\
\label{speed}&\left\| {{\mathbf{\dot q}}\left( t \right)} \right\| \le {V_{\max }},\forall t \in {{\mathcal{T}}},
\end{align}
\end{subequations}
\vspace{-1cm}

\noindent where the first derivative of ${{\mathbf{q}}}\left( t \right)$ with respect to $t$, ${{\mathbf{\dot q}}\left( t \right)}$, denotes the velocity vector, and $\overline \gamma$ denotes the minimum required expected effective channel power gain, which has to be achieved throughout the travel of the MRU. Constraints \eqref{Initial Location Constraint} and \eqref{speed} represent the mobility constraints on the MRU, where ${V_{\max }}$ is the maximum travelling speed. Problem \eqref{P_single} is challenging to solve for the following three reasons. Firstly, constraint \eqref{single user rate requirment2} is not concave with respect to ${\mathbf{q}}\left( t \right)$ and ${\mathbf{\Theta}}  \left( t \right)$. The unit modulus constraint \eqref{IRS phase} is also non-convex. Secondly, the expected effective channel power gain ${\lambda _m}\left( t \right)$ is generally not a continuous function under the considered location-dependent channel model. Thirdly, problem \eqref{P_single} involves an infinite number of optimization variables with respect to continuous time $t$, which are difficult to handle. To tackle these difficulties, we develop a radio map based approach which is capable of exploiting knowledge regarding location-dependent channel propagation.
\vspace{-0.6cm}
\subsection{Channel Power Gain Map Construction}
\vspace{-0.1cm}
In this subsection, we introduce a specific type of radio map, namely, the channel power gain map. Specifically, the channel power gain map characterizes the spatial distribution of the expected effective channel power gain over the region of interest with respect to the MRU's location ${\mathbf{q}}$, i.e., ${\lambda _m}\left( {\mathbf{q}} \right)$. \textcolor{black}{Let ${\overline X }$ and ${\overline Y }$ denote the range of the two-dimensional (2D) space along the x-axis and y-axis, respectively. For the development of the radio map, the continuous 2D  space is first discretized into $\frac{{\overline X}}{{{\Delta _x}}} \times \frac{{\overline Y}}{{{\Delta _y}}}$ small cells, where $\Delta_x$ and $\Delta_y$ denote the resolution of the radio map along the x-axis and y-axis, respectively, which should be chosen such that the location of the MRU within each cell can be approximated by the cell center. To guarantee a certain desired accuracy of the approximation, the values of $\Delta_x$ and $\Delta_y$ can be selected such that ${\Delta _x} \le \varepsilon \overline X $ and ${\Delta _y} \le \varepsilon \overline Y $, where $\varepsilon $ is a predefined threshold for the accuracy\footnote{\textcolor{black}{Though small values of $\varepsilon$ (i.e., small $\Delta_x$ and $\Delta_y$) increase the accuracy of the approximation, they also cause a high computational complexity for robot path planning due to the resulting large number of cells. In practice, $\varepsilon$ has to be properly chosen to balance between accuracy and complexity.}}. The horizontal location of the $\left( {i,j} \right)$-th cell center can be expressed as}
% The size of each cell is denoted by $\Delta ={\Delta _x}{\Delta _y}$.
\vspace{-0.5cm}
\textcolor{black}{\begin{align}\label{location u}
{\mathbf{q}}_{i,j}^\Delta  = {{\mathbf{q}}_0} + \left[ {\left( {i - 1} \right){\Delta _x},\left( {j - 1} \right){\Delta _y}} \right] ,i \in {\mathcal{X}},j \in {\mathcal{Y}},
\end{align}}
\vspace{-1.3cm}

\noindent \textcolor{black}{where ${{\mathbf{q}}_0}$ is the center of the cell in the lower left corner of the considered 2D space, ${\mathcal{X}} = \left\{ {1, \ldots ,X} \right\}$, ${\mathcal{Y}} = \left\{ {1, \ldots ,Y} \right\}$, $X \triangleq \frac{{\overline X }}{\Delta_x }$, and $Y \triangleq \frac{{\overline Y }}{\Delta_y }$.}\\
\indent Accordingly, let matrix ${\mathbf{C}} \in {{\mathbb{R}}^{X \times Y}}$ denote the channel power gain map, where the element in row $i$ and column $j$ characterizes the maximum expected effective channel power gain of the MRU at location $\left\{ {{\mathbf{q}}_{i,j}^\Delta } \right\}$. Therefore, the elements of ${\mathbf{C}}$ are given by
\vspace{-0.3cm}
\begin{align}\label{channel gain map}
{\left[ {\mathbf{C}} \right]_{i,j}} = \mathop {\max }\limits_{{\mathbf{\Theta }} \in {{\mathcal{F}}}} {\left| {\widetilde h_m^H\left( {{\mathbf{q}}_{i,j}^\Delta } \right) + \widetilde {\mathbf{w}}_m^H\left( {{\mathbf{q}}_{i,j}^\Delta } \right){\bm{\theta }\left( {{\mathbf{q}}_{i,j}^\Delta } \right)}} \right|^2} + {\tau _m}\left( {{\mathbf{q}}_{i,j}^\Delta } \right),i \in {{\mathcal{X}}},j \in {{\mathcal{Y}}},
\end{align}
\vspace{-1cm}

\noindent  where ${\mathcal{F}}$ denotes the set of all possible IRS reflection matrices.\\
\indent For any given ${\mathbf{q}}_{i,j}^\Delta$, the expected effective channel power gain is upper-bounded by
\vspace{-0.3cm}
\begin{align}\label{upper bound}
{\left| {\widetilde h_m^H\left( {{\mathbf{q}}_{i,j}^\Delta } \right) + \widetilde {\mathbf{w}}_m^H\left( {{\mathbf{q}}_{i,j}^\Delta } \right){\bm{\theta }}} \right|^2} + {\tau _m}\left( {{\mathbf{q}}_{i,j}^\Delta } \right) \le {\left( {\left| {\widetilde h_m^H\left( {{\mathbf{q}}_{i,j}^\Delta } \right)} \right| + {{\left\| {\widetilde {\mathbf{w}}_m^H\left( {{\mathbf{q}}_{i,j}^\Delta } \right)} \right\|}_1}} \right)^2} + {\tau _m}\left( {{\mathbf{q}}_{i,j}^\Delta } \right).
\end{align}
\vspace{-1cm}

\noindent The above inequality holds with equality for the following optimal phase shifts:
\vspace{-0.3cm}
\begin{align}\label{ap theta}
\theta _n^*\left( {{\mathbf{q}}_{i,j}^\Delta } \right) = \angle \left( {\widetilde h_m^H{{\left( {{\mathbf{q}}_{i,j}^\Delta } \right)}}} \right) - \angle \left( {{{\left[ {\widetilde {\mathbf{w}}_m^H\left( {{\mathbf{q}}_{i,j}^\Delta } \right)} \right]}_n}} \right),\forall n \in {{\mathcal{N}}},
\end{align}
\vspace{-1cm}

\noindent where ${\angle \left(  \cdot   \right)}$ denotes the phase of a complex number. Therefore, the channel power gain map ${\mathbf{C}}$ is given as follows:
\begin{figure}[b!]
    \begin{center}
        \includegraphics[width=2.4in]{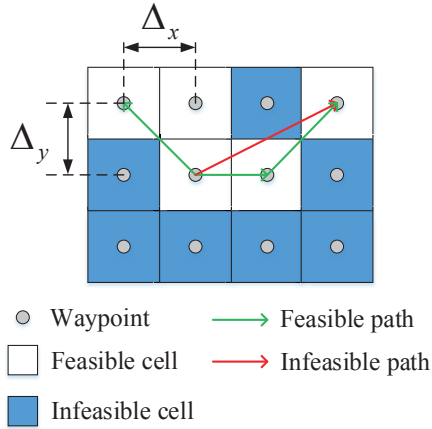}
        \caption{\textcolor{black}{Illustration of path structure for problem \eqref{graph single}.}}
        \label{path structure}
    \end{center}
\end{figure}
\vspace{-0.3cm}
\begin{align}\label{obtained channel gain map}
  {\left[ {\mathbf{C}} \right]_{i,j}} = {\left( {\left| {\widetilde h_m^H\left( {{\mathbf{q}}_{i,j}^\Delta } \right)} \right| + {{\left\| {\widetilde {\mathbf{w}}_m^H\left( {{\mathbf{q}}_{i,j}^\Delta } \right)} \right\|}_1}} \right)^2} + {\tau _m}\left( {{\mathbf{q}}_{i,j}^\Delta } \right), i \in {\mathcal{X}},j \in {\mathcal{Y}}.
\end{align}
\vspace{-1.6cm}
\subsection{Graph Theory Based Path Solution}
\vspace{-0.0cm}
Based on the constructed channel power gain map, let ${{Q}} \!=\! \left\{\! {{\mathbf{q}}_{{i_1},{j_1}}^\Delta ,{\mathbf{q}}_{{i_2},{j_2}}^\Delta , \ldots ,{\mathbf{q}}_{{i_{D - 1}},{j_{D - 1}}}^\Delta ,{\mathbf{q}}_{{i_D},{j_D}}^\Delta \!} \right\}$ denote the path of the MRU. For ease of exposition, we assume that ${\mathbf{q}}_{{i_1},{j_1}}^\Delta  = {{\mathbf{q}}_I}$ and ${\mathbf{q}}_{{i_D},{j_D}}^\Delta  = {{\mathbf{q}}_F}$. It can be verified that for the optimal solution of \eqref{P_single}, the speed constraint \eqref{speed} must be satisfied with equality, i.e., $\left\| {{\mathbf{\dot q}}\left( t \right)} \right\| = {V_{\max }},\forall t \in {{\mathcal{T}}}$. To demonstrate this, suppose that for the optimal solution to problem \eqref{P_single}, the MRU travels at a speed strictly less than ${V_{\max }}$. Then, we can increase the speed to ${V_{\max }}$, which decreases the travelling time. With this insight, problem \eqref{P_single} can be equivalently reformulated as the following \emph{travelling distance minimization} problem over the channel power gain map:
\vspace{-0.3cm}
\begin{subequations}\label{graph single}
\begin{align}
&\mathop {\min }\limits_{{{Q}},D} \sum\limits_{d = 1}^{D - 1} {\left\| {{\mathbf{q}}_{{i_{d + 1}},{j_{d + 1}}}^\Delta  - {\mathbf{q}}_{{i_d},{j_d}}^\Delta } \right\|}\\
\label{graph single QoS}{\rm{s.t.}}\;\;& {\left[ {\mathbf{C}} \right]_{{i_d},{j_d}}} \ge {\overline \gamma} ,\\
\label{adjacent single}&\left\| {{\mathbf{q}}_{{i_{d + 1}},{j_{d + 1}}}^\Delta  - {\mathbf{q}}_{{i_d},{j_d}}^\Delta } \right\| \le \textcolor{black}{\sqrt {\Delta _x^2 + \Delta _y^2}}  ,1 \le d \le D - 1,\\
\label{graph single Initial Location}&{\mathbf{q}}_{{i_1},{j_1}}^\Delta  = {{\mathbf{q}}_I},{\mathbf{q}}_{{i_D},{j_D}}^\Delta  = {{\mathbf{q}}_F},
\end{align}
\end{subequations}
\vspace{-1cm}

\noindent where \eqref{adjacent single} ensures that any two successive waypoints along the path are adjacent in the channel power gain map. As illustrated in Fig. \ref{path structure}, if the two successive waypoints satisfy the expected effective channel power gain condition, it is guaranteed that any point on the line segment connecting them also satisfies this condition (e.g., green lines). Otherwise, if the two successive waypoints were not adjacent, the path may not be always feasible (e.g., the red line). However, problem \eqref{graph single} is a non-convex combinatorial optimization problem, which is difficult to solve with standard convex optimization methods. In the following, we solve problem \eqref{graph single} by exploiting graph theory \cite{Graph}.\\
\indent For given ${\overline \gamma}$ and channel power gain map ${\mathbf{C}}$, we construct a new matrix ${\mathbf{\Pi}}  \in {{\mathbb{R}}^{X \times Y}}$, namely the feasible map, as follows:
\vspace{-0.3cm}
\begin{align}\label{feasible map}
{\left[ {\mathbf{\Pi}}  \right]_{i,j}}  = \left\{ \begin{gathered}
  1,\;\;{\rm{if}}\;{\left[ {\mathbf{C}} \right]_{i,j}} \ge \;{\overline \gamma}  \hfill \\
  0,\;{\rm{otherwise}} \hfill \\
\end{gathered}  \right.,i \in {\mathcal{X}},j \in {\mathcal{Y}}.
\end{align}
\vspace{-0.8cm}

\noindent Specifically, ${\left[ {\mathbf{\Pi}}  \right]_{i,j}} = 1$ means that the location ${{\mathbf{q}}_{i,j}^\Delta }$ is a feasible candidate waypoint for the path of the MRU.\\
\indent Based on the feasible map ${\mathbf{\Pi}}$, we construct an undirected weighted graph, which is denoted by $G = \left( {V,E} \right)$. The vertex set $V$ and the edge set $E$ are given by
\vspace{-0.3cm}
\begin{subequations}
\begin{align}\label{V}&V = \left\{ {{{\mathbf{v}}_{i,j}} = {\mathbf{q}}_{i,j}^\Delta :{{\left[ {\mathbf{\Pi}}  \right]}_{i,j}} = 1,i \in {\mathcal{X}},j \in {\mathcal{Y}}} \right\},\\
\label{E}&E = \left\{ {\left( {{{\mathbf{v}}_{i,j}},{{\mathbf{v}}_{i',j'}}} \right):{{\mathbf{v}}_{i,j}},{{\mathbf{v}}_{i',j'}} \in V} \right\}.
\end{align}
\end{subequations}
\vspace{-1cm}

\noindent The weight of each edge is denoted by $W\left( {{{\mathbf{v}}_{i,j}},{{\mathbf{v}}_{i',j'}}} \right)$, and given by
\vspace{-0.3cm}
\begin{align}
W\left( {{{\mathbf{v}}_{i,j}},{{\mathbf{v}}_{i',j'}}} \right) = \left\{ \begin{gathered}
  \left\| {{{\mathbf{v}}_{i,j}} - {{\mathbf{v}}_{i',j'}}} \right\|,{\rm{if}}\left\| {{{\mathbf{v}}_{i,j}} - {{\mathbf{v}}_{i',j'}}} \right\| \le \textcolor{black}{\sqrt {\Delta _x^2 + \Delta _y^2}}\hfill \\
  \infty ,{\rm{otherwise}} \hfill \\
\end{gathered}.  \right.
\end{align}
\vspace{-0.8cm}

\indent Based on the constructed graph $G$, problem \eqref{graph single} is equivalent to finding the shortest path from ${\mathbf{v}}_{{i_1},{j_1}} = {{\mathbf{q}}_I}$ to ${\mathbf{v}}_{{i_D},{j_D}} = {{\mathbf{q}}_F}$. The shortest path construction problem can be efficiently solved via the Dijkstra algorithm \cite{Graph} with complexity ${\mathcal{O}}\left( {{{\left| V \right|}^2}} \right)$. The optimal path for the MRU is denoted by ${{{Q}}^*} = \left\{ {{\mathbf{q}}_{{i_1},{j_1}}^{*\Delta },{\mathbf{q}}_{{i_2},{j_2}}^{*\Delta }, \ldots ,{\mathbf{q}}_{{i_{D - 1}},{j_{D - 1}}}^{*\Delta },{\mathbf{q}}_{{i_D},{j_D}}^{*\Delta }} \right\}$.
\vspace{-0.3cm}
\section{Radio Map based Robot Path Planning for Multiple-user System}
\vspace{-0.1cm}
\subsection{System Model}
\vspace{-0.1cm}
\begin{figure}[h!]
    \begin{center}
        \includegraphics[width=2.4in]{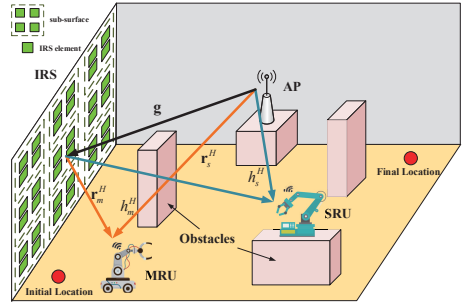}
        \caption{Illustration of the IRS-enhanced indoor robot navigation system for a multiple-user scenario.}
        \label{System model2}
    \end{center}
\end{figure}
\vspace{-0.0cm}
Different from the single-user scenario, where the MRU is assigned a dedicated resource block, we further investigate the scenario where multiple robotic users are simultaneously served by the AP. More particularly, two types of robotic users are considered, namely, an MRU and an SRU, as shown in Fig. \ref{System model2}\footnote{One practical scenario for the multiple-user setup is automatic industrial production in smart factories, where the MRU is used for transportation of materials, and the SRU is used for assembling products. Both users need to maintain a reliable communication link to ensure safe operation.}. The location of the SRU is denoted by ${\mathbf{u}_s}=\left( {x_s,y_s,{H_s}} \right)$. \textcolor{black}{Since the location of the SRU is usually selected to avoid signal blockage and to avoid collisions with the MRU (e.g., by placing the SRU on a high work platform),} Rician fading is assumed for the AP-SRU and IRS-SRU channels, respectively, which are modelled as follows:
\vspace{-0.3cm}
\begin{align}\label{AP-static}
{h_s} = \frac{{\sqrt{{{\mathcal{L}}}_{AS}}}}{{\sqrt {{K_{AS}} + 1} }}\left( {\sqrt {{K_{AS}}} {{\overline h}_s} + {{\widehat h}_s}} \right),
\end{align}
\vspace{-1cm}
\begin{align}\label{IRS-static}
{{\mathbf{r}}_s} = \frac{{\sqrt{{{\mathcal{L}}}_{IS}}}}{{\sqrt {{K_{IS}} + 1} }}\left( {\sqrt {{K_{IS}}} {{\overline {\mathbf{r}} }_s} + {{\widehat {\mathbf{r}}}_s}} \right),
\end{align}
\vspace{-1cm}

\noindent where ${{{{\mathcal{L}}}_{AS}}}$ and ${{{{\mathcal{L}}}_{IS}}}$ denote the distance-dependent path losses, ${\overline {{h}} _s}$ and ${\overline {\mathbf{r}} _s}$ denote the deterministic LoS components, and ${{\widehat h}_s}$ and ${\widehat {\mathbf{r}}_s}$ denote the random NLoS components. $K_{AS}$ and $K_{IS}$ are the Rician factors.\\
\indent Therefore, the effective channel from the AP to the SRU is given by ${c_s}\left( t \right) = h_s^H + {\mathbf{r}}_s^H{\mathbf{\Theta}} \left( t \right){\mathbf{g}}$. Similar to \textbf{Lemma 1} and \eqref{expected robotic channel gain}, the expected effective channel power gain of the SRU can be expressed as
\vspace{-0.3cm}
\begin{align}\label{expected static channel gain}
\begin{gathered}
  {\mathbb{E}}\left[ {{{\left| {{c_s}\left( t \right)} \right|}^2}} \right] = {\left| {\widetilde h_s^H + \widetilde {\mathbf{r}}_s^H{\mathbf{\Theta }}\left( t \right)\widetilde {\mathbf{g}}} \right|^2} + \frac{{{{{\mathcal{L}}}_{AS}}}}{{{K_{AS}} + 1}} + \frac{{{{{\mathcal{L}}}_{AI}}{{{\mathcal{L}}}_{IS}}\left( {{K_{IS}} + {K_{AI}} + 1} \right)M}}{{\left( {{K_{AI}} + 1} \right)\left( {{K_{IS}} + 1} \right)}} \hfill \\
  \;\;\;\;\;\;\;\;\;\;\;\;\;\;\;\;\;\; = {\left| {\widetilde h_s^H + \widetilde {\mathbf{w}}_s^H{\bm{\theta }}\left( t \right)} \right|^2} + {\tau _s} \triangleq {\lambda _s}\left( t \right), \hfill \\
\end{gathered}
\end{align}
\vspace{-0.6cm}

\noindent where $\widetilde h_s^H = \sqrt {\frac{{{{{\mathcal{L}}}_{AS}}}{{K_{AS}}}}{{{K_{AS}} + 1}}} \overline h _s^H$, ${\widetilde {\mathbf{r}}_s^H} = \sqrt {\frac{{{{{\mathcal{L}}}_{IS}}}{{K_{IS}}}}{{{K_{IS}} + 1}}} {\overline {\mathbf{r}} _s^H}$, ${\mathbf{w}}_s^H = \widetilde {\mathbf{r}}_s^H{\rm{diag}}\left( {\widetilde {\mathbf{g}}} \right) \in {\mathbb{C}^{1 \times M}}$, ${\left[ {{{\widetilde {\mathbf{w}}}_s^H}}\in {\mathbb{C}^{1 \times N}} \right]_n} = \sum\nolimits_{\overline n = 1}^{\overline N} {{{\left[ {{{\mathbf{w}}_s^H}} \right]}_{\overline n + \left( {n - 1} \right)\overline N}}} ,\forall n \in {{\mathcal{N}}}$, and ${\tau _s} = \frac{{{{{\mathcal{L}}}_{AS}}}}{{{K_{AS}} + 1}} + \frac{{{{{\mathcal{L}}}_{AI}}{{{\mathcal{L}}}_{IS}}\left( {{K_{IS}} + {K_{AI}} + 1} \right)M}}{{\left( {{K_{AI}} + 1} \right)\left( {{K_{IS}} + 1} \right)}}$.\\
\indent Regarding the multiple access scheme applied at the AP for serving the two robotic users, both NOMA and OMA transmission are considered. In NOMA, the AP simultaneously serves the MRU and the SRU in the same time/frequency resource blocks by utilizing superposition coding (SC) and successive interference cancelation (SIC)\footnote{\textcolor{black}{SC and SIC required for NOMA introduce additional complexity compared to OMA. However, on the other hand, NOMA yields a significant performance gain over OMA, see Section IV-B for details.}}. For OMA, we focus on frequency division multiple access (FDMA), where the AP simultaneously serves the two users in different frequency resource blocks\footnote{For time division multiple access (TDMA), the AP needs to serve the two users consecutively in different time resource blocks, which causes transmission delays. Therefore, we consider FDMA to ensure a fair comparison with NOMA.}.
\subsubsection{NOMA} According to the NOMA protocol, the AP transmits the two users' signals using superposition coding. The received signal of user $l \in \left\{ {s,m} \right\}$ at time instant $t$ can be expressed as
\vspace{-0.6cm}
\begin{align}\label{received signal NOMA}
{e_l}\left( t \right) = {{c_l}\left( t \right)} \left( {\sqrt {{p_s}\left( t \right)} {b_s}\left( t \right) + \sqrt {{p_m}\left( t \right)} {b_m}\left( t \right)} \right) + {n_l}\left( t \right),
\end{align}
\vspace{-1.2cm}

\noindent where ${{b_s}\left( t \right)}$ and ${{b_m}\left( t \right)}$ are the transmitted data symbols for the SRU and the MRU, respectively, which are modelled as circularly symmetric complex Gaussian (CSCG) random variables with zero mean and unit variance. Let $P_{\max}$ denote the maximum transmit power at the AP. The power allocation of the two users has to satisfy ${p_s}\left( t \right) + {p_m}\left( t \right) \le {P_{\max }},\forall t$. ${n_l}\left( t \right)$ denotes the additive CSCG noise at user $l \in \left\{ {s,m} \right\}$ with zero mean and variance ${\sigma ^2}$.\\
\indent By invoking SIC, the user with the stronger channel power gain is able to first decode the signal of the user with the weaker channel power gain, before decoding its own signal \cite{Liu2017}. We define binary indicators ${\mu _l}\left( t \right) \in \left\{ {0,1} \right\},l \in \left\{ {s,m} \right\}$, to specify the instantaneous decoding order of the two users, which satisfy ${\mu _{m}}\left( t \right) + {\mu _{s}}\left( t \right) = 1,\forall t$. For instance, if the MRU is the strong user, we have ${\mu _{m}}\left( t \right)=0$ and ${\mu _{s}}\left( t \right)=1$. In this case, the effective channel power gain should satisfy ${\left| {{c_m}\left( t \right)} \right|^2} \ge {\left| {{c_s}\left( t \right)} \right|^2}$ to ensure the success of SIC \cite{Liu2017}. Therefore, the achievable communication rate of user $l \in \left\{ {s,m} \right\}$ can be expressed as
\vspace{-0.3cm}
\begin{align}\label{rate NOMA}
  R_l^{\rm{NOMA}}\left( t \right) = {\log _2}\left( {1 + \frac{{{{\left| {{c_l}\left( t \right)} \right|}^2}{p_l}\left( t \right)}}{{{\mu _l}\left( t \right){{\left| {{c_l}\left( t \right)} \right|}^2}{p_{\overline l }}\left( t \right) + {\sigma ^2}}}} \right) = {\log _2}\left( {1 + \frac{{{p_l}\left( t \right)}}{{{\mu _l}\left( t \right){p_{\overline l }}\left( t \right) + \frac{{{\sigma ^2}}}{{{{\left| {{c_l}\left( t \right)} \right|}^2}}}}}} \right).
\end{align}
\vspace{-0.8cm}

\noindent Here, if $l = m$, we have $\overline l  = s$; otherwise, $\overline l  = m$. Note that $R_l^{\rm{NOMA}}\left( t \right)$ is a random variable, and we are interested in the expected/average achievable communication rate, defined as ${\mathbb{E}}\left[ {R_l^{\rm{NOMA}}\left( t \right)} \right]$. However, it is difficult to derive a closed-form expression for ${\mathbb{E}}\left[ {R_l^{\rm{NOMA}}\left( t \right)} \right]$, since its probability distribution is hard to obtain. To tackle this issue, we approximate the expected achievable communication rate by an upper bound as follows:
\vspace{-0.3cm}
\begin{align}\label{approximated rate NOMA}
  {\mathbb{E}}\left[ {R_l^{\rm{NOMA}}\left( t \right)} \right] \! \mathop  \le \limits^{\left( a \right)}\! {\log _2}\!\left( {1\! + \! \frac{{{p_l}\left( t \right)}}{{{\mu _l}\left( t \right){p_{\overline l }}\left( t \right) + \frac{{{\sigma ^2}}}{{{\mathbb{E}}\left[ {{{\left| {{c_l}\left( t \right)} \right|}^2}} \right]}}}}} \right) \!\!=\! {\log _2}\!\left( {1\! +\! \frac{{{p_l}\left( t \right)}}{{{\mu _l}\left( t \right){p_{\overline l }}\left( t \right) + \frac{{{\sigma ^2}}}{{{\lambda _l}\left( t \right)}}}}} \right)\!\! \triangleq \!\overline R _l^{\rm{NOMA}}\left( t \right),
\end{align}
\vspace{-0.8cm}

\noindent where ${\left( a \right)}$ holds due to the Jensen's inequality since the rate function $R_l^{\rm{NOMA}}\left( t \right)$ is concave with respect to ${{{\left| {{c_l}\left( t \right)} \right|}^2}}$ \textcolor{black}{and $\left\{ {{p_l}\left( t \right)} \right\}$ are deterministic variables which are selected by the AP, see Section III-B.} The tightness of the approximation $\overline R _l^{\rm{NOMA}}\left( t \right)$ with respect to the exact average rate ${\mathbb{E}}\left[ {R_l^{\rm{NOMA}}\left( t \right)} \right]$ will be evaluated in Section VI-B4.
\subsubsection{OMA} For OMA transmission, the AP simultaneously transmits to both users in orthogonal frequency bands of equal size. Accordingly, the achievable communication rate for user $l \in \left\{ {s,m} \right\}$ is given by
\vspace{-0.5cm}
\begin{align}\label{rate OMA}
R_l^{\rm{OMA}}\left( t \right) = \frac{1}{2}{\log _2}\left( {1 + \frac{{{{\left| {{c_l}\left( t \right)} \right|}^2}{p_l}\left( t \right)}}{{\frac{1}{2}{\sigma ^2}}}} \right).
\end{align}
\vspace{-1.0cm}

\noindent Similarly, the expected achievable communication rate for OMA can be approximated as
\vspace{-0.3cm}
\begin{align}\label{approximated rate OMA}
  {\mathbb{E}}\left[ {R_l^{\rm{OMA}}\left( t \right)} \right] \le \frac{1}{2}{\log _2}\left( {1 + \frac{{{\mathbb{E}}\left[ {{{\left| {{c_l}\left( t \right)} \right|}^2}} \right]{p_l}\left( t \right)}}{{\frac{1}{2}{\sigma ^2}}}} \right) = \frac{1}{2}{\log _2}\left( {1 + \frac{{2{\lambda _l}\left( t \right){p_l}\left( t \right)}}{{{\sigma ^2}}}} \right) \triangleq \overline R _l^{\rm{OMA}}\left( t \right),
\end{align}
\vspace{-2cm}
\subsection{Problem Formulation}
\vspace{-0.3cm}
\indent For the multiple-user scenario, the communication-aware robot path planning problem is formulated as follows:
\vspace{-0.6cm}
\begin{subequations}\label{PNOMA}
\begin{align}
&\mathop {\min }\limits_{Q,P,A,T} \;T\\
\label{communication rate requirement NOMA}{\rm{s.t.}}\;\;& {\overline R _l^Z}\left( t \right) \ge {\overline r_l}, Z \in \left\{ {{\rm{NOMA}},{\rm{OMA}}} \right\}, l \in \left\{ {s,m} \right\},\forall t \in {{\mathcal{T}}},\\
\label{power allocation NOMA}&{p_l}\left( t \right) \ge 0,l \in \left\{ {s,m} \right\},{p_m}\left( t \right) + {p_s}\left( t \right) \le {P_{\max }},\forall t \in {{\mathcal{T}}},\\
\label{decoding order NOMA1}&{\mu _l}\left( t \right) \in \left\{ {0,1} \right\},l \in \left\{ {s,m} \right\},\forall t \in {{\mathcal{T}}},\\
\label{decoding order NOMA2}&{\mu _{m}}\left( t \right) + {\mu _{s}}\left( t \right) = 1,\forall t \in {{\mathcal{T}}},\\
\label{decoding order NOMA3}&\left\{ \begin{gathered}
  {\lambda _s}\left( t \right) \ge {\lambda _m}\left( t \right),{\rm{if}}\;{\mu _s}\left( t \right) = 0,{\mu _m}\left( t \right) = 1 \hfill \\
  {\lambda _s}\left( t \right) \le {\lambda _m}\left( t \right),{\rm{otherwise}} \hfill \\
\end{gathered}  \right.,\forall t \in {{\mathcal{T}}},\\
\label{CONX}&\eqref{IRS phase}-\eqref{speed},
\end{align}
\end{subequations}
\vspace{-1.2cm}

\noindent where $P = \left\{ {{p_m}\left( t \right),{p_s}\left( t \right),0 \le t \le T} \right\}$ denotes the power allocation at the AP, ${\overline r_l},l \in \left\{ {s,m} \right\}$, denote the minimum required communication rate of MRU and SRU, and $Z \in \left\{ {{\rm{NOMA}},{\rm{OMA}}} \right\}$ indicates whether NOMA or OMA is employed. Constraint \eqref{power allocation NOMA} denotes the power allocation constraint. Constraints \eqref{decoding order NOMA1}-\eqref{decoding order NOMA3} ensure that SIC can be successfully implemented at the stronger user and is only valid when NOMA is used, i.e., $Z={\rm{NOMA}}$.
\vspace{-0.6cm}
\subsection{Communication Rate Map Construction}
\vspace{-0.1cm}
Different from the single-user scenario, where the communication performance of the MRU is only determined by its location and the IRS phase shifts, the communication performance in problem \eqref{PNOMA} is also determined by the power allocation at the AP. Moreover, the effective channels of the MRU and the SRU share the same IRS phase shifts, which makes problem \eqref{PNOMA} challenging to solve. To tackle this difficulty, we construct a different type of radio map, which we refer to as the \emph{communication rate map} of the MRU, by jointly optimizing the power allocation at the AP and the phase shifts at the IRS.
\subsubsection{NOMA} Let matrix ${\mathbf{U}}^{\rm{NOMA}} \in {{\mathbb{R}}^{X \times Y}}$ denote the communication rate map for NOMA. The elements of ${\mathbf{U}}^{\rm{NOMA}}$ represent the maximum expected communication rates of the MRU at locations $\left\{ {{\mathbf{q}}_{i,j}^\Delta } \right\}$, such that the communication rate requirement of the SRU is satisfied. Therefore, the element in row $i$ and column $j$ of ${\mathbf{U}}^{\rm{NOMA}}$ is given by
\vspace{-0.3cm}
\begin{subequations}\label{rate map NOMA}
\begin{align}
&{\left[ {\mathbf{U}}^{\rm{NOMA}} \right]_{i,j}} =\mathop {\max }\limits_{{p_m},{p_s},{\mathbf{\Theta}} ,{\mu _m},{\mu _s},{\overline r_0}} \;\overline r_0\\
\label{static rate requirement NOMA}{\rm{s.t.}}\;\;& {\overline R _m^{\rm{NOMA}}}\left( {{\mathbf{q}}_{i,j}^\Delta } \right) \ge {\overline r_0},\\
\label{static rate requirement NOMA2}& {\overline R _s^{\rm{NOMA}}}\left( {{\mathbf{q}}_{i,j}^\Delta } \right) \ge {\overline r_s},\\
\label{rate map NOMA IRS phase}&{\theta _n}\in \left[ {0,2\pi } \right),\forall n \in {{\mathcal{N}}},\\
\label{rate map NOMA power allocation NOMA}&{p_l} \ge 0,l \in \left\{ {s,m} \right\},{p_m} + {p_s} \le {P_{\max }},\\
\label{rate map decoding order NOMA1}&{\mu _l} \in \left\{ {0,1} \right\},l \in \left\{ {s,m} \right\},\\
\label{rate map decoding order NOMA2}&{\mu _{m}} + {\mu _{s}} = 1,\\
\label{rate map decoding order NOMA3}&\left\{ \begin{gathered}
  {\lambda _s} \ge {\lambda _m}\left( {{\mathbf{q}}_{i,j}^\Delta } \right),{\rm{if}}\;{\mu _s} = 0,{\mu _m} = 1 \hfill \\
  {\lambda _s} \le {\lambda _m}\left( {{\mathbf{q}}_{i,j}^\Delta } \right),{\rm{otherwise}} \hfill \\
\end{gathered}  \right.,
\end{align}
\end{subequations}
\vspace{-0.8cm}

\noindent where $\overline r_0$ denotes the maximum expected communication rate of the MRU. Let ${{\overline r}_0^*}$ denote the optimal solution of problem \eqref{rate map NOMA}. As there are $2!=2$ options for the decoding order for two users, we can solve problem \eqref{rate map NOMA} by exhaustively searching over all possible decoding order options, i.e., ${{\overline r}_0^*} = \mathop {\arg \max }\limits_{k \in \left\{ {1,2} \right\}} \left( {{{\overline r}_0^{k*}}} \right)$, where ${{{\overline r}_0^{k*}} }$ denotes the optimal solution of problem \eqref{rate map NOMA} for the $k$th decoding order option.\\
\indent To solve problem \eqref{rate map NOMA}, we first introduce auxiliary variables $\widetilde {\mathbf{h}}_m^H = \left[ {\widetilde {\mathbf{w}}_s^H\left( {{\mathbf{q}}_{i,j}^\Delta } \right)\;\widetilde h_m^H\left( {{\mathbf{q}}_{i,j}^\Delta } \right)} \right]$, $\widetilde {\mathbf{h}}_s^H = \left[ {\widetilde {\mathbf{w}}_s^H\;\widetilde h_s^H} \right]$, and ${\mathbf{v}} = {\left[ {{e^{j{\theta _1}}}, {e^{j{\theta _2}}}, \ldots, {e^{j{\theta _N}}}, 1} \right]^T}$. Moreover, we define ${\widetilde {\mathbf{H}}_l} = {\widetilde {\mathbf{h}}_l}\widetilde {\mathbf{h}}_l^H, l \in \left\{ {s,m} \right\}$, and ${\mathbf{V}} = {\mathbf{v}}{{\mathbf{v}}^H}$, which satisfies ${{\mathbf{V}}} \succeq 0$, ${\rm {rank}}\left( {{{\mathbf{V}}}} \right) = 1$, and ${\left[ {\mathbf{V}} \right]_{nn}} = 1,n = 1,2, \ldots ,N + 1$. Then, the expected effective channel power gain of the MRU and the SRU can be rewritten as
\vspace{-0.5cm}
\begin{align}
{\lambda _m}\left( {{\mathbf{q}}_{i,j}^\Delta } \right) ={\left| {\widetilde {\mathbf{h}}_m^H{\mathbf{v}}} \right|^2} + {\tau _m}\left( {{\mathbf{q}}_{i,j}^\Delta } \right) = {\rm{Tr}}\left( {{{\widetilde {\mathbf{H}}}_m}{\mathbf{V}}} \right) + {\tau _m}\left( {{\mathbf{q}}_{i,j}^\Delta } \right),
\end{align}
\vspace{-1.2cm}
\begin{align}
{\lambda _s} = {\left| {\widetilde {\mathbf{h}}_s^H{\mathbf{v}}} \right|^2} + {\tau _s} = {\rm{Tr}}\left( {{{\widetilde {\mathbf{H}}}_s}{\mathbf{V}}} \right) + {\tau _s}.
\end{align}
\vspace{-1.0cm}

\noindent For a given user decoding order\footnote{Here, we consider the case where the SRU is the strong user, i.e., ${\mu _s} = 0,{\mu _m} = 1$. A similar problem can be also formulated for ${\mu _s} = 1,{\mu _m} = 0$.}, problem \eqref{rate map NOMA} can be reformulated as
\vspace{-0.3cm}
\begin{subequations}\label{rate map NOMA order}
\begin{align}
&\mathop {\max }\limits_{{p_m},{p_s},{\mathbf{V}},{\overline r_0^k}} \;\overline r_0^k\\
\label{static rate requirement NOMA order}{\rm{s.t.}}\;\;& {\log _2}\left( {1 + \frac{{{p_m}}}{{{p_s} + \frac{{{\sigma ^2}}}{{{\rm{Tr}}\left( {{{\widetilde {\mathbf{H}}}_m}{\mathbf{V}}} \right) + {\tau _m}\left( {{\mathbf{q}}_{i,j}^\Delta } \right)}}}}} \right) \ge {\overline r _0^k},\\
\label{static rate requirement NOMA2 order}& {\log _2}\left( {1 + \frac{{{p_s}\left( {{\rm{Tr}}\left( {{{\widetilde {\mathbf{H}}}_s}{\mathbf{V}}} \right) + {\tau _s}} \right)}}{{{\sigma ^2}}}} \right) \ge {\overline r _s},\\
\label{SIC}& {\rm{Tr}}\left( {{{\widetilde {\mathbf{H}}}_s}{\mathbf{V}}} \right) + {\tau _s} \ge {\rm{Tr}}\left( {{{\widetilde {\mathbf{H}}}_m}{\mathbf{V}}} \right) + {\tau _m}\left( {{\mathbf{q}}_{i,j}^\Delta } \right),\\
\label{Vmm}&{\left[ {\mathbf{V}} \right]_{nn}} = 1,n = 1,2, \ldots ,N + 1,\\
\label{SDP V}&{{\mathbf{V}}}  \succeq  0, {\mathbf{V}} \in {{\mathbb{H}}^{N + 1}},\\
\label{rank 1 V}&{\rm {rank}}\left( {{{\mathbf{V}}}} \right) = 1,\\
\label{IRS phase NOMAij2 order}&\eqref{rate map NOMA power allocation NOMA}.
\end{align}
\end{subequations}
\vspace{-1.2cm}

\noindent Due to the non-convex constraints \eqref{static rate requirement NOMA order}, \eqref{static rate requirement NOMA2 order}, and \eqref{rank 1 V}, problem \eqref{rate map NOMA order} is a non-convex optimization problem, and hence, difficult to solve globally optimally. To address this issue, we develop an efficient bisection search based algorithm to derive a high-quality suboptimal solution. First, for a given rate target $\overline r _0^t$, the non-convex constraints \eqref{static rate requirement NOMA order} and \eqref{static rate requirement NOMA2 order} can be rearranged as
\vspace{-0.3cm}
\begin{align}\label{c1}
{p_m} \ge \left( {{2^{\overline r _0^t}} - 1} \right)\left( {{p_s} + \frac{{{\sigma ^2}}}{{{\rm{Tr}}\left( {{{\widetilde {\mathbf{H}}}_m}{\mathbf{V}}} \right) + {\tau _m}\left( {{\mathbf{q}}_{i,j}^\Delta } \right)}}} \right),
\end{align}
\vspace{-0.8cm}
\begin{align}\label{c2}
{\rm{Tr}}\left( {{{\widetilde {\mathbf{H}}}_s}{\mathbf{V}}} \right) + {\tau _s} \ge \frac{{\left( {{2^{{{\overline r }_s}}} - 1} \right){\sigma ^2}}}{{{p_s}}}.
\end{align}
\vspace{-1.0cm}

\noindent Then, we have the following feasibility check problem:
\vspace{-0.6cm}
\begin{subequations}\label{feasible check}
\begin{align}
&\mathop {\max }\limits_{{p_m},{p_s},{\mathbf{V}}} \;1\\
{\rm{s.t.}}\;\;&\eqref{rate map NOMA power allocation NOMA},\eqref{SIC}-\eqref{rank 1 V},\eqref{c1},\eqref{c2}.
\end{align}
\end{subequations}
\vspace{-1.2cm}

\noindent For a given rate target $\overline r _0^t$, if problem \eqref{feasible check} is feasible, it follows that ${{{\overline r}_0^{k*}}} \ge \overline r_0^t$, otherwise, ${{{\overline r}_0^{k*}}} < \overline r_0^t$. Therefore, problem \eqref{rate map NOMA order} can be solved by successively checking the feasibility of problem \eqref{feasible check} with updated $\overline r_0^t$'s until the bisection search terminates. However, problem \eqref{feasible check} is non-convex due to the non-convex rank-one constraint \eqref{rank 1 V}. To handle this difficulty, we first transform rank constraint \eqref{rank 1 V} equivalently into the follow constraint:
\vspace{-0.6cm}
\begin{align}\label{dc}
{\left\| {\mathbf{V}} \right\|_ * } - {\left\| {\mathbf{V}} \right\|_2} \le 0,
\end{align}
\vspace{-1.2cm}

\noindent where ${\left\| {\mathbf{V}} \right\|_ * } = \sum\nolimits_i {{\sigma _i}\left( {\mathbf{V}} \right)}$ and ${\left\| {\mathbf{V}} \right\|_2} = {\sigma _1}\left( {\mathbf{V}} \right)$ denote the nuclear norm and spectral norm, respectively, and ${\sigma _i}\left( {\mathbf{V}} \right)$ is the $i$th largest singular value of matrix ${\mathbf{V}}$. For any ${\mathbf{V}} \in {{\mathbb{H}}^{N + 1}}$, we have ${\left\| {\mathbf{V}} \right\|_ * } - {\left\| {\mathbf{V}} \right\|_2} \ge 0$ and equality holds if and only if ${\mathbf{V}}$ is a rank-one matrix. Therefore, the feasibility of problem \eqref{feasible check} can be checked by solving the following problem:
\begin{subequations}\label{feasible check2}
\vspace{-0.6cm}
\begin{align}
&\mathop {\min }\limits_{{p_m},{p_s},{\mathbf{V}}} {\left\| {\mathbf{V}} \right\|_ * } - {\left\| {\mathbf{V}} \right\|_2}\\\
{\rm{s.t.}}\;\;&\eqref{rate map NOMA power allocation NOMA},\eqref{SIC}-\eqref{SDP V},\eqref{c1},\eqref{c2}.
\end{align}
\end{subequations}
\vspace{-1.2cm}

\noindent Specifically, if the objective function of problem \eqref{feasible check2} is zero, it means that a rank-one solution can be obtained and problem \eqref{feasible check} is feasible, otherwise, problem \eqref{feasible check} is infeasible. However, problem \eqref{feasible check2} is still non-convex due to the non-convex objective function. In the following, we invoke SCA~\cite{SCA} to find a suboptimal solution of \eqref{feasible check2} iteratively.\\
\indent As the objective function of \eqref{feasible check2} is a difference of convex (DC) functions, for a given feasible point ${{{\mathbf{V}}^{\widetilde n}}}$ in the ${\widetilde n}$th iteration of the SCA method, a lower bound on ${\left\| {\mathbf{V}} \right\|_2}$ is constructed via a first-order Taylor expansion as follows:
\vspace{-0.6cm}
\begin{align}\label{lower bound}
{\left\| {\mathbf{V}} \right\|_2} \ge {\left\| {{{\mathbf{V}}^{\widetilde n}}} \right\|_2} + {\rm{Tr}}\left[ {{{\mathbf{u}}_{\max }}\left( {{{\mathbf{V}}^{\widetilde n}}} \right){{\left( {{{\mathbf{u }}_{\max }}\left( {{{\mathbf{V}}^{\widetilde n}}} \right)} \right)}^H}\left( {{\mathbf{V}} - {{\mathbf{V}}^{\widetilde n}}} \right)} \right] \triangleq {\overline {\mathbf{V}} ^{\widetilde n}},
\end{align}
\vspace{-1.2cm}

\noindent where ${{{\mathbf{u}}_{\max }}\left( {{{\mathbf{V}}^{\widetilde n}}} \right)}$ denotes the eigenvector corresponding to the largest eigenvalue of ${{{\mathbf{V}}^{\widetilde n}}}$.\\
\indent In the ${\widetilde n}$th iteration for a given feasible point, ${{{\mathbf{V}}^{\widetilde n}}}$, by replacing $\left\| {\mathbf{V}} \right\|$ with its lower bound ${\overline {\mathbf{V}} ^{\widetilde n}}$, we can find an upper bound on problem \eqref{feasible check2} by solving the following optimization problem:
\vspace{-0.6cm}
\begin{subequations}\label{feasible check3}
\begin{align}
&\mathop {\min }\limits_{{p_m},{p_s},{\mathbf{V}}} {\left\| {\mathbf{V}} \right\|_ * } - {\overline {\mathbf{V}} ^{\widetilde n}}\\\
{\rm{s.t.}}\;\;&\eqref{rate map NOMA power allocation NOMA},\eqref{SIC}-\eqref{SDP V},\eqref{c1},\eqref{c2}.
\end{align}
\end{subequations}
\vspace{-1.2cm}

\noindent Note that problem \eqref{feasible check3} is a convex semidefinite program (SDP), which can be efficiently solved by existing convex optimization solvers such as CVX \cite{cvx}. The proposed SCA based algorithm for solving problem \eqref{feasible check2} is summarized in \textbf{Algorithm 1}, where the matrix solution obtained in a given iteration is used as the feasible point for the next iteration. By iteratively solving problem \eqref{feasible check3}, the objective function of \eqref{feasible check3} is monotonically non-increasing and the proposed \textbf{Algorithm 1} is guaranteed to converge to a stationary point of \eqref{feasible check2}.\\
\begin{algorithm}[!t]\label{method1}
\caption{SCA based Algorithm for Problem \eqref{feasible check2}}
\begin{algorithmic}[1]
\STATE {Initialize ${{\mathbf{V}}^0}$, and set iteration index $\widetilde n=0$.}
\STATE {\bf repeat}
\STATE \quad Solve problem \eqref{feasible check3} for given ${{\mathbf{V}}^{\widetilde n}}$.
\STATE \quad Update ${{\mathbf{V}}^{\widetilde n+1}}$ with the obtained optimal solution, and $\widetilde n=\widetilde n+1$.
\STATE {\bf until convergence}
\STATE  Return the converged objective function of \eqref{feasible check3}.
\end{algorithmic}
\end{algorithm}
\begin{algorithm}[!t]\label{method2}
\caption{Bisection Search based Algorithm for Determining the Elements of ${\mathbf{U}}^{\rm{NOMA}}$}
\hspace*{0.02in}
\hspace*{0.02in} {Given $\left\{ {{\mathbf{q}}_{i,j}^\Delta } \right\}$.}\\
\vspace{-0.4cm}
\begin{algorithmic}[1]
\STATE {\bf for} $k=1,2$ {\bf do}
\STATE \quad Initialize $r_{\max}$, $r_{\min}$, and the $k$th decoding order option ${\mu _s},{\mu _m}$. Set the defined accuracy $\varepsilon_0 $ of the bisection search.
\STATE \quad {\bf while} ${r_{\max }} - {r_{\min }} \ge \varepsilon_0 $, {\bf do}
\STATE \quad\quad Solve problem \eqref{feasible check} without the rank-one constraint for given $\overline r _0^t = \frac{{{r_{\max }} + {r_{\min }}}}{2}$.
\STATE \qquad\quad{\bf if} the relaxed version of \eqref{feasible check} is unsolvable, {\bf then}
\STATE \qquad\quad\quad Problem \eqref{feasible check} is infeasible, ${r_{\max }} = \overline r _0^t$.
\STATE \qquad\quad{\bf else}
\STATE \qquad\quad\quad Denote the optimal solution of the relaxed problem by ${{\mathbf{V}}^*}$.
\STATE \qquad\quad\quad Solve problem \eqref{feasible check2} by applying \textbf{Algorithm 1} with ${{\mathbf{V}}^0} = {{\mathbf{V}}^*}$.
\STATE \qquad\qquad {\bf if} the converged objective function is zero, {\bf then}
\STATE \qquad\qquad\quad Problem \eqref{feasible check} is feasible, ${r_{\min }} = \overline r _0^t$.
\STATE \qquad\qquad{\bf else}
\STATE \qquad\qquad\quad Problem \eqref{feasible check} is infeasible, ${r_{\max }} = \overline r _0^t$.
\STATE \qquad\qquad{\bf end if}
\STATE \qquad\quad{\bf end if}
\STATE \quad{\bf end while}
\STATE \quad${{{\overline r}_0^{k*}}} = {r_{\min }}$.
\STATE {\bf end for}
\STATE  ${{\overline r}_0^*} = \mathop {\arg \max }\limits_{k \in \left\{ {1,2} \right\}} \left( {{{\overline r}_0^{k*}}} \right)$.
\end{algorithmic}
\end{algorithm}
\indent The overall bisection search based algorithm for determining the elements of ${\mathbf{U}}^{\rm{NOMA}}$ is summarized in \textbf{Algorithm 2}, where \textbf{Algorithm 1} is applied to check the feasibility of problem \eqref{feasible check} for a given rate target $\overline r _0^t$. Note that in \textbf{Algorithm 1}, a feasible matrix ${{{\mathbf{V}}^0}}$, which does not have to be rank-one, has to be initialized. To find such a matrix, we first solve problem \eqref{feasible check} by applying semidefinite relaxation (SDR) and ignoring the rank-one constraint. The relaxed version of \eqref{feasible check} can be efficiently solved by existing convex optimization solvers such as CVX \cite{cvx}. We note that if the relaxed version of \eqref{feasible check} is unsolvable, this means that problem \eqref{feasible check} is also infeasible for the rate target $\overline r _0^t$. In this case, we do not have to apply the proposed \textbf{Algorithm 1}, and can directly enter the next iteration of the bisection search algorithm by updating the current upper bound of the rate target as $\overline r _0^t$. If the relaxed version of \eqref{feasible check} is solvable, we initialize ${{{\mathbf{V}}^0}}$ in \textbf{Algorithm 1} with the optimal solution of the relaxed problem, denoted by ${{\mathbf{V}}^*}$, and check the feasibility of problem \eqref{feasible check} based on the result obtained from \textbf{Algorithm 1}. Furthermore, according to \cite{Luo}, the complexities of solving the relaxed version of \eqref{feasible check} and applying \textbf{Algorithm 1} are ${{\mathcal{O}}}\left( {{{\left( {N + 1} \right)}^{4.5}}} \right)$ and ${{\mathcal{O}}}\left( {{I}{{\left( {N + 1} \right)}^{4.5}}} \right)$, respectively, where ${{I}}$ denotes the number of iterations needed for convergence of \textbf{Algorithm 1}. Thus, the overall complexity of \textbf{Algorithm 2} with two user decoding order options is ${{\mathcal{O}}}\left( 2{{{\log }_2}\left( {\frac{{{r_{\max }}-{r_{\min }}}}{\varepsilon_0 }} \right)\left( {\left( {{I} + 1} \right){{\left( {N + 1} \right)}^{4.5}}} \right)} \right)$, where ${{{r }_{\max }}}$ and ${{{r }_{\min }}}$ are the initial upper and lower bounds of the bisection search, respectively, and $\varepsilon_0$ denotes the accuracy of the bisection search.
%, which in general determines the final converged solution. In order to obtain a high quality solution,
%Note that a feasible but unnecessary rank-one matrix  for solving problem \eqref{feasible check3}. To achieve this, we first solve problem \eqref{feasible check} by ignoring the rank-one constraint to obtain the initial matrix ${{{\mathbf{V}}^0}}$. Then, problem \eqref{feasible check3} is solved in an iterative manner until convergence.  It is worth noting that if problem \eqref{feasible check} without the rank-one constraint is infeasible, problem \eqref{feasible check3} is also infeasible. The proposed bisection search based algorithm for determining the elements of ${\mathbf{U}}^{\rm{NOMA}}$ is summarized in \textbf{Algorithm 1}.
\subsubsection{OMA} Let matrix ${\mathbf{U}}^{\rm{OMA}} \in {{\mathbb{R}}^{X \times Y}}$ denote the communication rate map for OMA. The element in row $i$ and column $j$ of ${\mathbf{U}}^{\rm{OMA}}$ can be obtained by solving the following problem:
\vspace{-0.3cm}
\begin{subequations}\label{rate map OMA}
\begin{align}
&{\left[ {\mathbf{U}}^{\rm{OMA}} \right]_{i,j}} =\mathop {\max }\limits_{{p_m},{p_s},{\mathbf{\Theta}} ,{\overline r_0}} \;\overline r_0\\
\label{static rate requirement OMA}{\rm{s.t.}}\;\;& {\overline R _m^{\rm{OMA}}}\left( {{\mathbf{q}}_{i,j}^\Delta } \right) \ge {\overline r_0},\\
\label{static rate requirement OMA2}& {\overline R _s^{\rm{OMA}}}\left( {{\mathbf{q}}_{i,j}^\Delta } \right) \ge {\overline r_s},\\
\label{rate map OMA IRS phase}&\eqref{rate map NOMA IRS phase},\eqref{rate map NOMA power allocation NOMA}.
\end{align}
\end{subequations}
\vspace{-1.2cm}

\noindent With the auxiliary variables introduced in the previous subsection, problem \eqref{rate map OMA} can be reformulated as follows:
\vspace{-0.6cm}
\begin{subequations}\label{rate map OMA check}
\begin{align}
&{\left[ {\mathbf{U}}^{\rm{OMA}} \right]_{i,j}} =\mathop {\max }\limits_{{p_m},{p_s},{\mathbf{V}} ,{\overline r_0}} \;\overline r_0\\
\label{static rate requirement OMA check}{\rm{s.t.}}\;\;& {\rm{Tr}}\left( {{{\widetilde {\mathbf{H}}}_m}{\mathbf{V}}} \right) + {\tau _m}\left( {{\mathbf{q}}_{i,j}^\Delta } \right) \ge \frac{{\left( {{2^{2{{\overline r }_0}}} - 1} \right){\sigma ^2}}}{{2{p_m}}},\\
\label{static rate requirement OMA2 check}& {\rm{Tr}}\left( {{{\widetilde {\mathbf{H}}}_s}{\mathbf{V}}} \right) + {\tau _s} \ge \frac{{\left( {{2^{2{{\overline r }_s}}} - 1} \right){\sigma ^2}}}{{2{p_s}}},\\
\label{rate map NOMA power allocation NOMA check}&\eqref{rate map NOMA power allocation NOMA},\eqref{Vmm},\eqref{SDP V},\eqref{rank 1 V}.
\end{align}
\end{subequations}
\vspace{-1.2cm}

\noindent As can be observed, problem \eqref{rate map OMA check} has a similar structure as problem \eqref{rate map NOMA order} in the previous subsection. Therefore, problem \eqref{rate map OMA check} can also be efficiently solved by the proposed bisection search based algorithm in \textbf{Algorithm 2} with the complexity of ${{\mathcal{O}}}\left( {{{\log }_2}\left( {\frac{{{r_{\max }}-{r_{\min }}}}{\varepsilon_0 }} \right)\left( {\left( {{I} + 1} \right){{\left( {N + 1} \right)}^{4.5}}} \right)} \right)$, where now only one decoding order has to be considered since OMA does not employ SIC.
\vspace{-0.6cm}
\subsection{Graph Theory based Path Solution}
\vspace{-0.1cm}
With the obtained communication rate map ${{\mathbf{U}}^Z},Z \in \left\{ {{\rm{NOMA}},{\rm{OMA}}} \right\}$, problem \eqref{PNOMA} is reformulated as follows:
\vspace{-0.5cm}
\begin{subequations}
\begin{align}\label{graph NOMA}
&\mathop {\min }\limits_{{{Q}},D} \sum\limits_{d = 1}^{D - 1} {\left\| {{\mathbf{q}}_{{i_{d + 1}},{j_{d + 1}}}^\Delta  - {\mathbf{q}}_{{i_d},{j_d}}^\Delta } \right\|}\\
\label{graph NOMA QoS}{\rm{s.t.}}\;\;& {\left[ {\mathbf{U}}^Z \right]_{{i_d},{j_d}}} \ge {\overline r_m}, Z \in \left\{ {{\rm{NOMA}},{\rm{OMA}}} \right\},\\
\label{graph NOMA constraints}&\eqref{adjacent single},\eqref{graph single Initial Location}.
\end{align}
\end{subequations}
\vspace{-1.2cm}

\noindent For given ${\overline r_m}$, we construct a feasible map based on ${\mathbf{U}}^Z$ as follows:
\vspace{-0.3cm}
\begin{align}\label{feasible map NOMA}
{\left[ {\mathbf{\Pi}}^Z  \right]_{i,j}}  = \left\{ \begin{gathered}
  1,\;\;{\rm{if}}\;{\left[ {\mathbf{U}}^Z\right]_{i,j}} \ge \;{\overline r_m}  \hfill \\
  0,\;{\rm{otherwise}} \hfill \\
\end{gathered}  \right.,i \in {\mathcal{X}},j \in {\mathcal{Y}}.
\end{align}
\vspace{-0.8cm}

\noindent To facilitate the application of graph theory, similar to the single-user case, we construct again an undirected weighted graph ${G^Z} = \left( {{V^Z},{E^Z}} \right)$, $Z \in \left\{ {{\rm{NOMA}},{\rm{OMA}}} \right\}$. Then, problem \eqref{PNOMA} can be solved by finding the shortest path from ${{\mathbf{q}}_I}$ to ${{\mathbf{q}}_F}$ in graph ${G^Z}$ via the Dijkstra algorithm. The details are omitted here for brevity.\\
\indent \textcolor{black}{Although we have considered a system with one MRU and one SRU so far in this section, the proposed radio map based path planning method can be easily extended to systems\footnote{\textcolor{black}{For systems with multiple MRUs, the robot path planning problem becomes more challenging since the multiple MRUs have to be carefully coordinated to avoid collisions and the communication rate map of each MRU depends on both its own location and the locations of the other MRUs. Therefore, this scenario is beyond the scope of this paper and constitutes an interesting topic for future work.}} with one MRU and multiple SRUs, which are indexed by ${s_k} \in \left\{ {{s_1},{s_2}, \ldots ,{s_{K}}} \right\} \triangleq {\mathcal{S}}$ with $K > 1$. In this case, the communication rate map characterization problem for both NOMA and OMA corresponds to finding the maximum expected communication rate of the MRU for each location, such that the communication rate requirements of the $K$ SRUs are satisfied. For NOMA, the resulting communication rate map characterization problem can also be solved with the proposed \textbf{Algorithm 2} by exhaustively searching over all $\left( {K + 1} \right)!$ possible NOMA user decoding orders. For OMA, the resulting communication rate map characterization problem can be solved with the proposed \textbf{Algorithm 2}, where the communication rate is calculated by allocating orthogonal frequency bands to the $K+1$ users. Therefore, the complexity of constructing the communication rate map for NOMA will significantly increase with the number of users due to the required exhaustive search over all possible user decoding orders, while the complexity of OMA does not grow\footnote{\textcolor{black}{We note that the communication rate map is constructed in an offline manner. Thus, the potentially high complexity of communication rate map computation for NOMA is acceptable given the available computing power.}}. Based on the communication rate map obtained for NOMA and OMA for systems with one MRU and multiple SRUs, the communication-aware robot path planning problem can be solved by employing graph theory.}
\vspace{-0.5cm}
\section{Numerical Examples}
\vspace{-0.2cm}
\begin{figure}[t!]
    \begin{center}
        \includegraphics[width=2.8in]{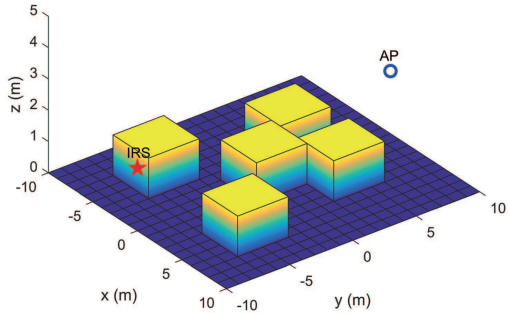}
        \caption{The simulated scenario (3D view).}
        \label{setup}
    \end{center}
\end{figure}
In this section, numerical examples are provided to validate the performance of the proposed IRS-enhanced robot navigation system. As illustrated in Fig. \ref{setup}, we consider an indoor factory (InF) environment with a width ${\overline X}$ and length ${\overline Y}$ of 20 meter, respectively, and a ceiling height of 5 meter. Specifically, the AP and the IRS are deployed at $\left( {0,10,2} \right)$ meters and $\left( {0,-10,2} \right)$ meters, respectively. The number of IRS reflecting elements in each sub-surface is set to $\overline N  = 20$. The total number of sub-surfaces is $N = {N_x}{N_z}$, where ${N_x}$ and ${N_z}$ denote the number of sub-surfaces along the $x$-axis and $z$-axis, respectively. Therefore, the total number of IRS reflecting elements is $M = \overline N {N_x}{N_z}$, where we set ${N_x}=10$ and increase ${N_z}$ linearly with $M$. The considered indoor environment includes 5 obstacles with a size of $4 \times 4 \times 1.3$ ${\rm{m}}^3$, respectively. The horizontal centers of the obstacles are located at $\left( { - 5, - 5} \right)$, $\left( { 5, - 5} \right)$, $\left( { 0, 0} \right)$, $\left( { - 3, 4} \right)$, and $\left( { 3, 4} \right)$ meters. The height of the antenna of the MRU is ${H_0} = 1$ m and its initial and final locations are ${{\mathbf{q}}_I} = \left( { - 10,0,1} \right)$ meters and ${{\mathbf{q}}_F} = \left( { 10,0,1} \right)$ meters, respectively. The path losses of all involved channels are modeled according to the 3rd Generation Partnership Project (3GPP) technical report for the InF-SH (sparse clutter, high BS) scenario \cite{3GPP}. For LoS channels, the path loss in dB is given by
\vspace{-0.6cm}
\begin{align}
{{\mathcal{L}}}_{{\rm{LoS}}} = 31.84 + 21.50{\log _{10}}\left( d \right) + 19{\log _{10}}\left({f_c}\right),
\end{align}
\vspace{-1.2cm}

\noindent where $d$ denotes the 3D distance between the robotic user and the AP (or the IRS), and ${f_c} = 2$ GHz is the carrier frequency. For NLoS channels, the path loss in dB is given by
\vspace{-0.6cm}
\begin{align}
{{\mathcal{L}}_{{\rm{NLoS}}}} = \max \left\{ {{{\mathcal{L}}_{{\rm{LoS}}}},32.4 + 23{{\log }_{10}}\left( d \right) + 20{{\log }_{10}}\left({f_c}\right)} \right\},
\end{align}
\vspace{-1.2cm}

\noindent which ensures that ${{\mathcal{L}}_{{\rm{NLoS}}}} \ge {{\mathcal{L}}_{{\rm{LoS}}}}$. The other system parameters are set as follows: The total transmit power of the AP is ${P_{\max }} = 20$ dBm, the noise power is ${\sigma ^2} =  - 90$ dBm, and the Rician factors of all involved channels are set to 3 dB. \textcolor{black}{To balance between the accuracy of approximating each cell by its center and the computational complexity of robot path planning, we set the threshold for the development of the radio map to $\varepsilon  = 0.025$. The corresponding resolutions are ${\Delta _x} = 0.5\;{\rm{m}} \le \varepsilon \overline X $ and ${\Delta _y} = 0.5\;{\rm{m}} \le \varepsilon \overline Y $.}
\vspace{-0.6cm}
\subsection{Single-user Scenario}
\vspace{-0.3cm}
In this subsection, we demonstrate the effectiveness of the proposed scheme for the single-user scenario. For comparison, we also consider the following benchmark schemes:
\begin{itemize}
  \item \textbf{IRS with discrete phase shifts}: In this case, the IRS is assumed to be equipped with finite resolution phase shifters. We have ${\theta _n} \in {\mathcal{D}} = \left\{ {0,\delta , \ldots ,\left( {L - 1} \right)\delta } \right\}$, where $\delta  = {{2\pi } \mathord{\left/
 {\vphantom {{2\pi } L}} \right.
 \kern-\nulldelimiterspace} L}$ and $L$ denotes the number of discrete phase shift levels. The corresponding channel power gain map is obtained by quantizing the optimal phase shift $\theta _n^* \left( {{\mathbf{q}}_{i,j}^\Delta } \right)$ in \eqref{ap theta} to the nearest discrete phase shift in ${\mathcal{D}}$ as follows:
 \vspace{-0.6cm}
 \begin{align}\label{ap1 theta}
{ \theta  _n^{\mathcal{D}}}\left( {{\mathbf{q}}_{i,j}^\Delta } \right) = \mathop {\arg \min  }\limits_{\theta  \in {\mathcal{D}}} \left| {\theta  - \theta _n^*\left( {{\mathbf{q}}_{i,j}^\Delta } \right)} \right|,\forall n \in {\mathcal{N}}.
\end{align}
\vspace{-1.2cm}
  \item \textbf{Without IRS}: In this case, the AP serves the user without the help of an IRS. The channel power gain map is obtained by considering only the AP-user channel.
\end{itemize}
\begin{figure}[b!]
\centering
\subfigure[Without IRS]{\label{withoutIRS}
\includegraphics[width= 2.5in]{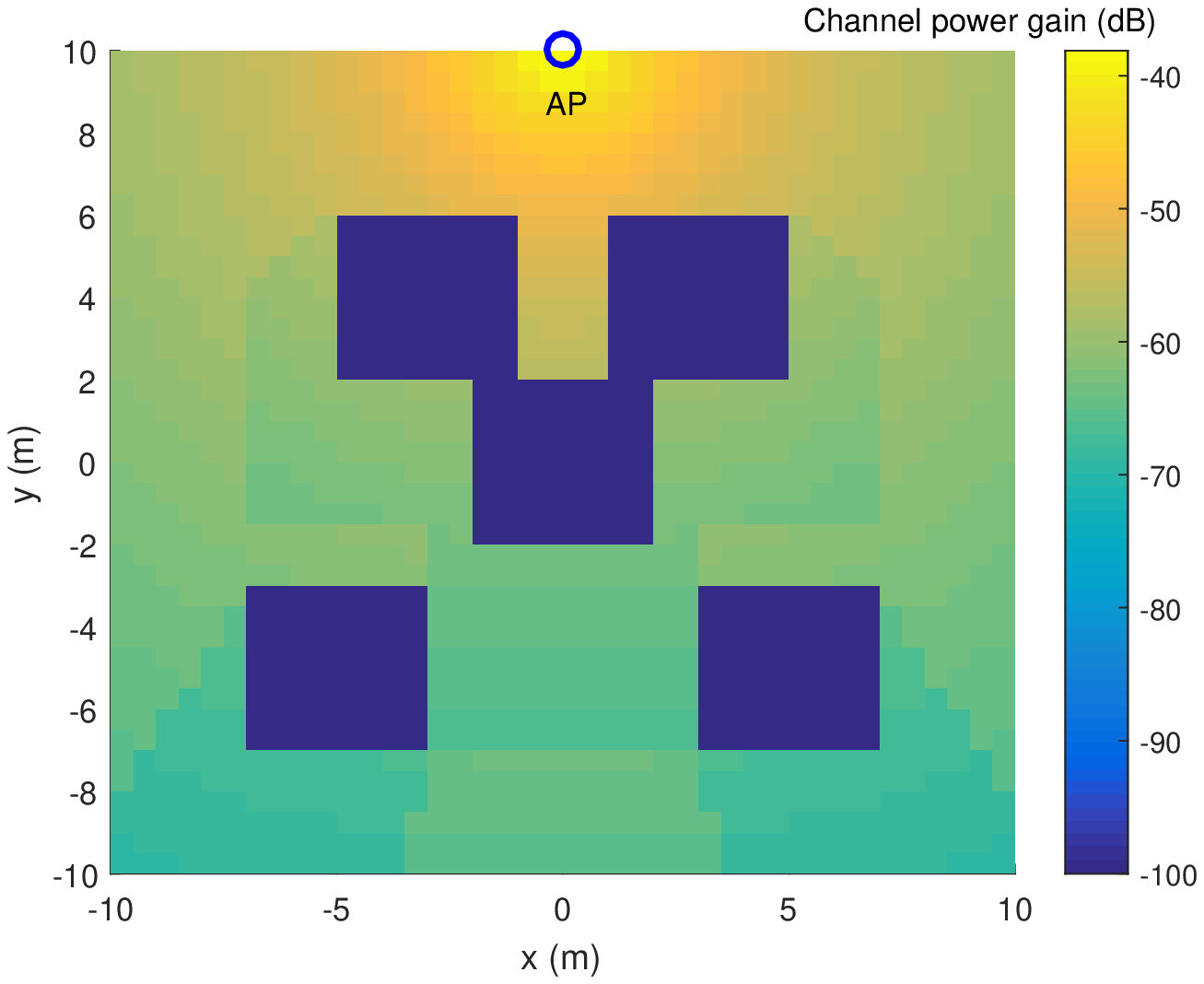}}
\subfigure[With IRS]{\label{withIRS}
\includegraphics[width= 2.5in]{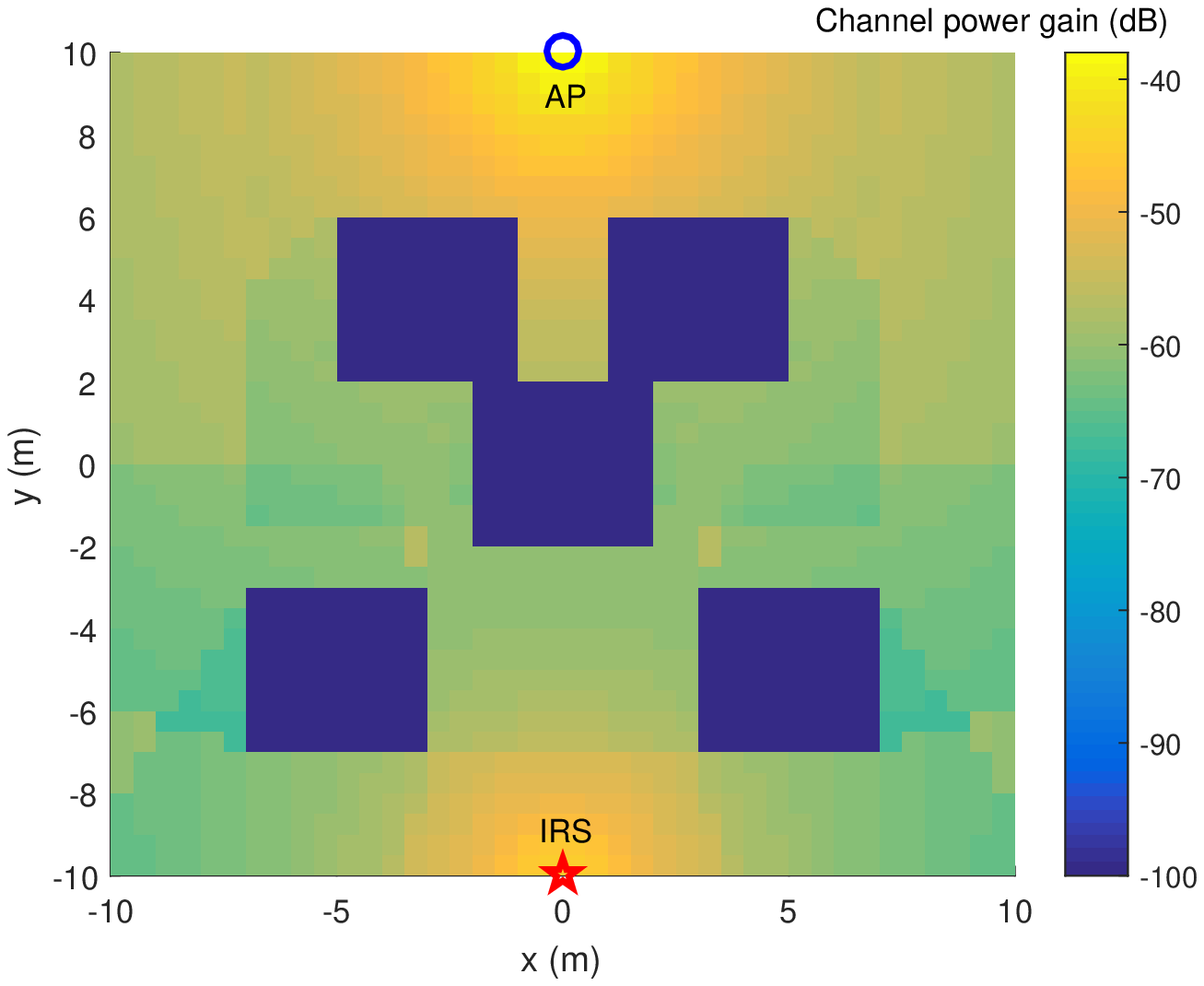}}
\setlength{\abovecaptionskip}{-0cm}
\caption{Illustration of the obtained channel power gain map for different schemes and \textcolor{black}{${\Delta_x }={\Delta_y }=0.5$ m}.}\label{channel gain map1}
\end{figure}
\subsubsection{Channel Power Gain Map} Fig. \ref{channel gain map1} illustrates the channel power gain map obtained from \eqref{channel gain map} with and without IRS, respectively. We set the number of reflecting elements to $M=1200$. As the MRU cannot enter the regions covered by obstacles, the corresponding expected channel power gain is set to $ - \infty $. One can observe that the distribution of the channel power gain changes abruptly due to the obstacles. Specifically, as depicted in Fig. \ref{withoutIRS}, without IRS, the channel power gains severely degrade if the AP-user link is blocked by obstacles. Moreover, from Fig. \ref{withIRS}, it can be observed that the channel power gains can be considerably improved by deploying an IRS, especially for the cells around the IRS. The IRS can be interpreted as a virtual AP, however, it is more energy-efficient than an actual AP since the IRS only passively reflects the incident signals.\\
\indent Based on the obtained channel power gain map, we investigate the percentage of cells, $\eta \left( {\overline \gamma  } \right)$, that can meet the expected channel power gain target, $\overline \gamma$. For a given $\overline \gamma$, $\eta \left( {\overline \gamma  } \right)$ is calculated as
\vspace{-0.3cm}
\begin{figure}[t!]
    \begin{center}
        \includegraphics[width=2.8in]{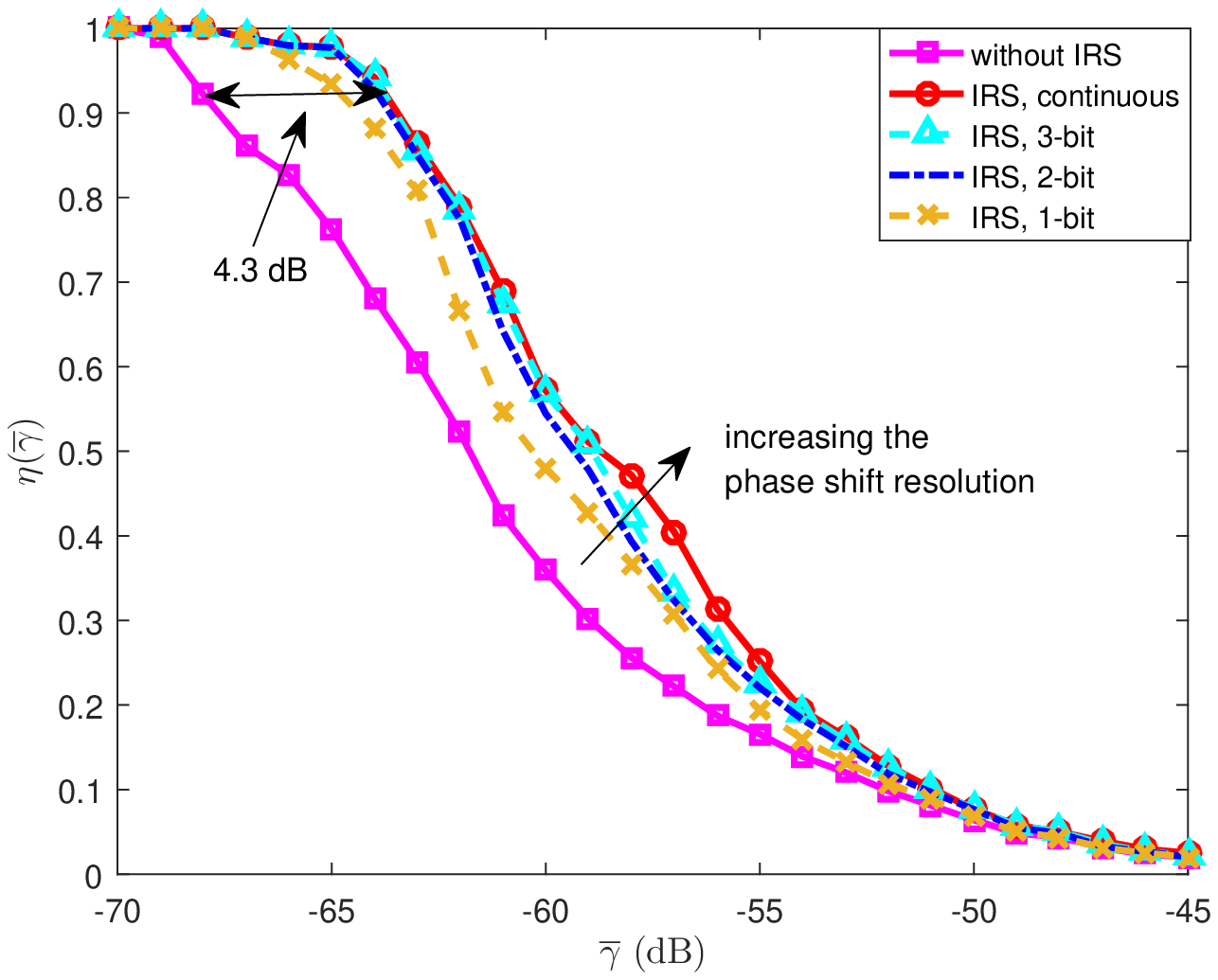}
        \caption{$\eta \left( {\overline \gamma  } \right)$ versus $\overline \gamma$ for different schemes.}
        \label{FR}
    \end{center}
\end{figure}
\begin{align}
\eta \left( {\overline \gamma  } \right) = \frac{{\sum\nolimits_{i = 1}^X {\sum\nolimits_{j = 1}^Y {{{\left[ {\mathbf{\Pi }} \right]}_{i,j}}} } }}{{XY - \Upsilon }},
\end{align}
\vspace{-1cm}

\noindent where $\Upsilon $ denotes the number of cells which are covered by obstacles. It is observed from Fig. \ref{FR} that $\eta \left( {\overline \gamma  } \right)$ decreases for both schemes as the expected channel power gain target increases. Specifically, without IRS, $\eta \left( {\overline \gamma  } \right)$ degrades more quickly than when the IRS is present. The proposed scheme with continuous phase shifts outperforms the scheme without IRS by up to 4.3 dB, which demonstrates the effectiveness of deploying IRSs to reduce the signal dead zones for indoor robotic communication. Moreover, for discrete phase shifts, 1-bit quantization leads to the worst performance as expected since only two phase shifts can be configured, which causes substantial performance loss. The performance achieved by discrete phase shifts approaches the upper bound achieved by continuous phase shifts as the phase shift resolution increases. For 2-bit and 3-bit quantization, the performance gap with respect to the continuous phase shift becomes negligible for most of the expected channel power gain targets, which suggests that 2- or 3-bit phase shifters are promising candidates for practical implementation.
\subsubsection{Obtained Paths of the MRU} 
\begin{figure*}[t!]
\centering
\subfigure[$\overline \gamma = -63$ dB]{\label{db63}
\includegraphics[width= 2in, height=1.6in]{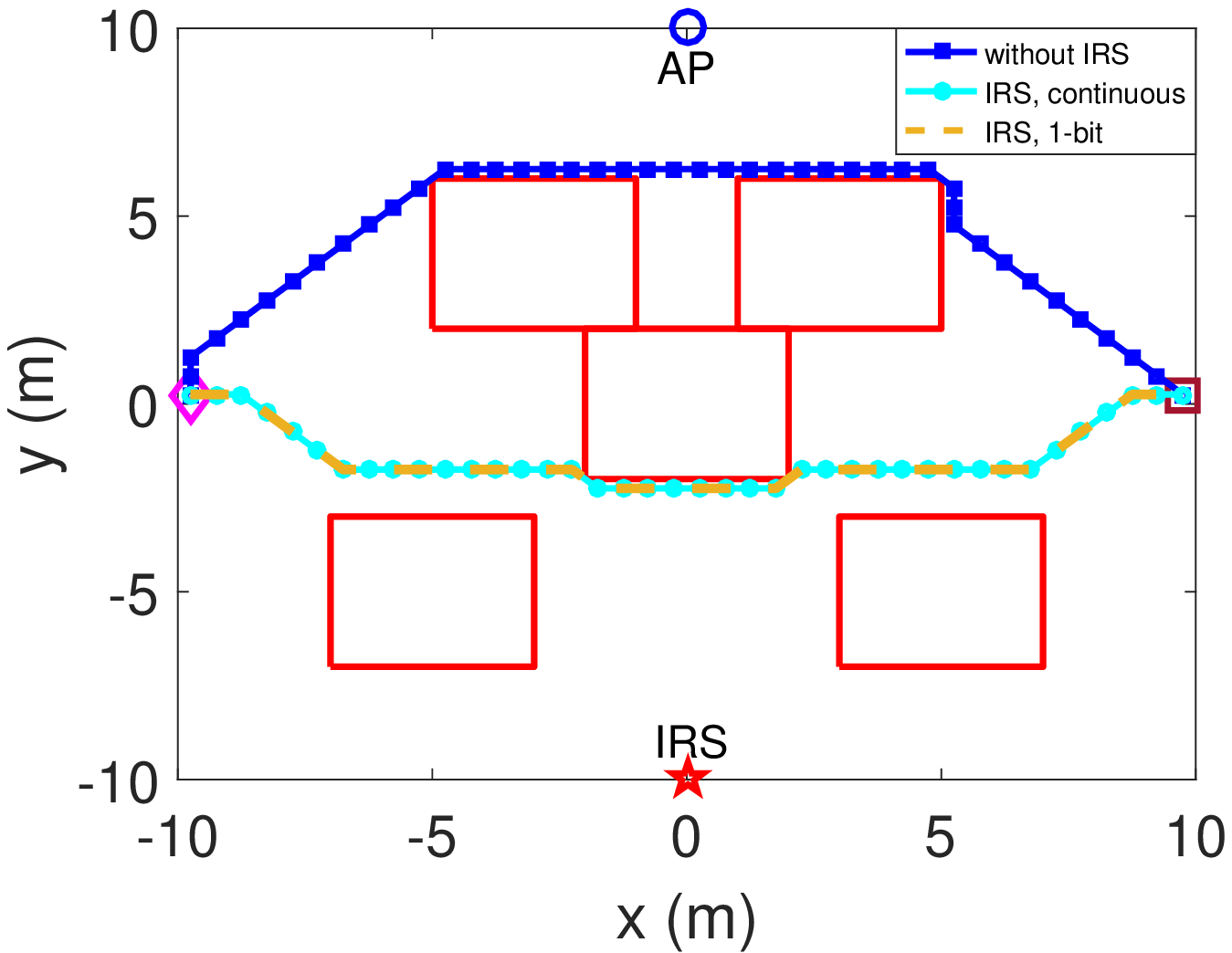}}
\subfigure[$\overline \gamma = -62.1$ dB]{\label{db62}
\includegraphics[width= 2in, height=1.6in]{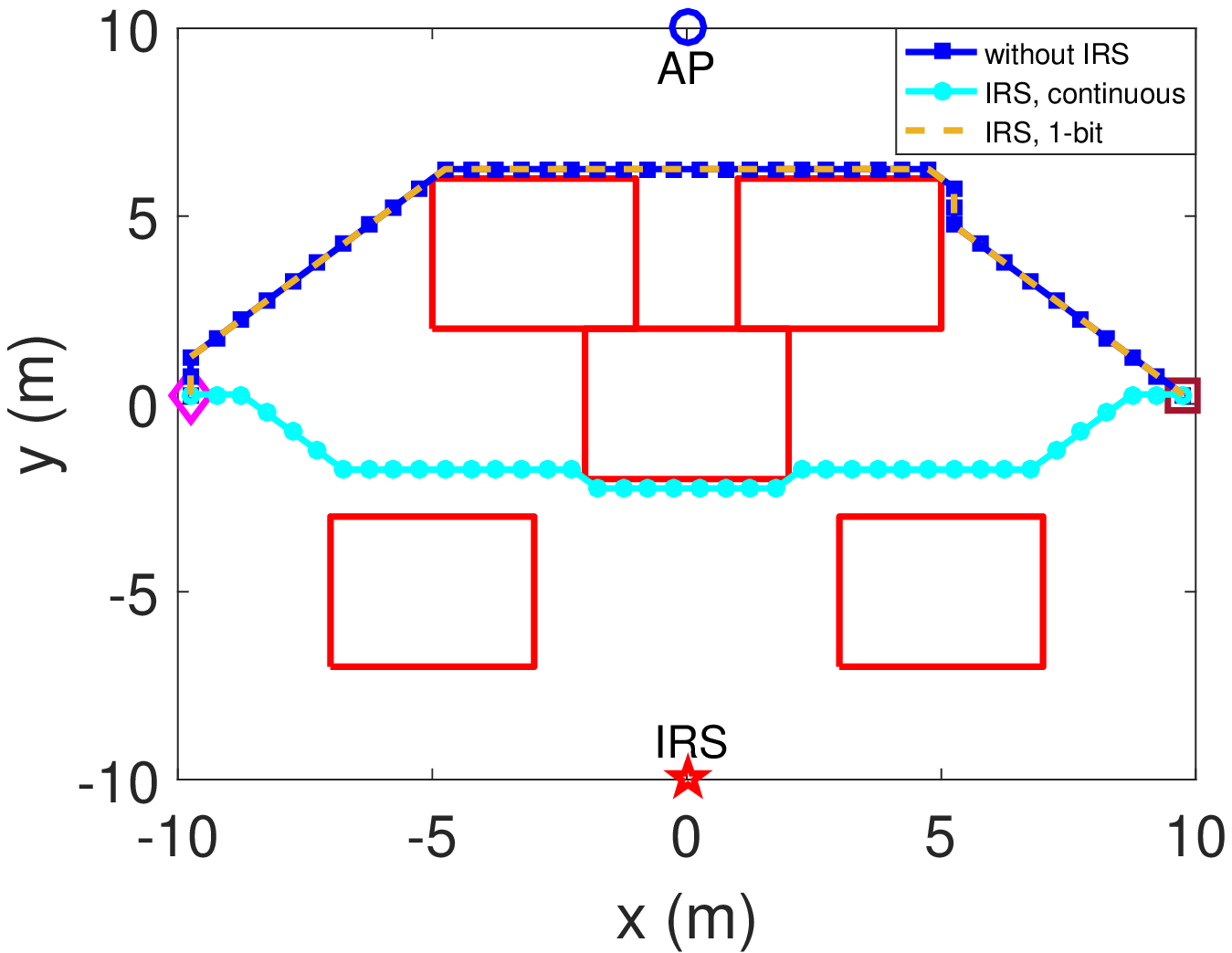}}
\subfigure[$\overline \gamma = -61$ dB]{\label{db61}
\includegraphics[width= 2in, height=1.6in]{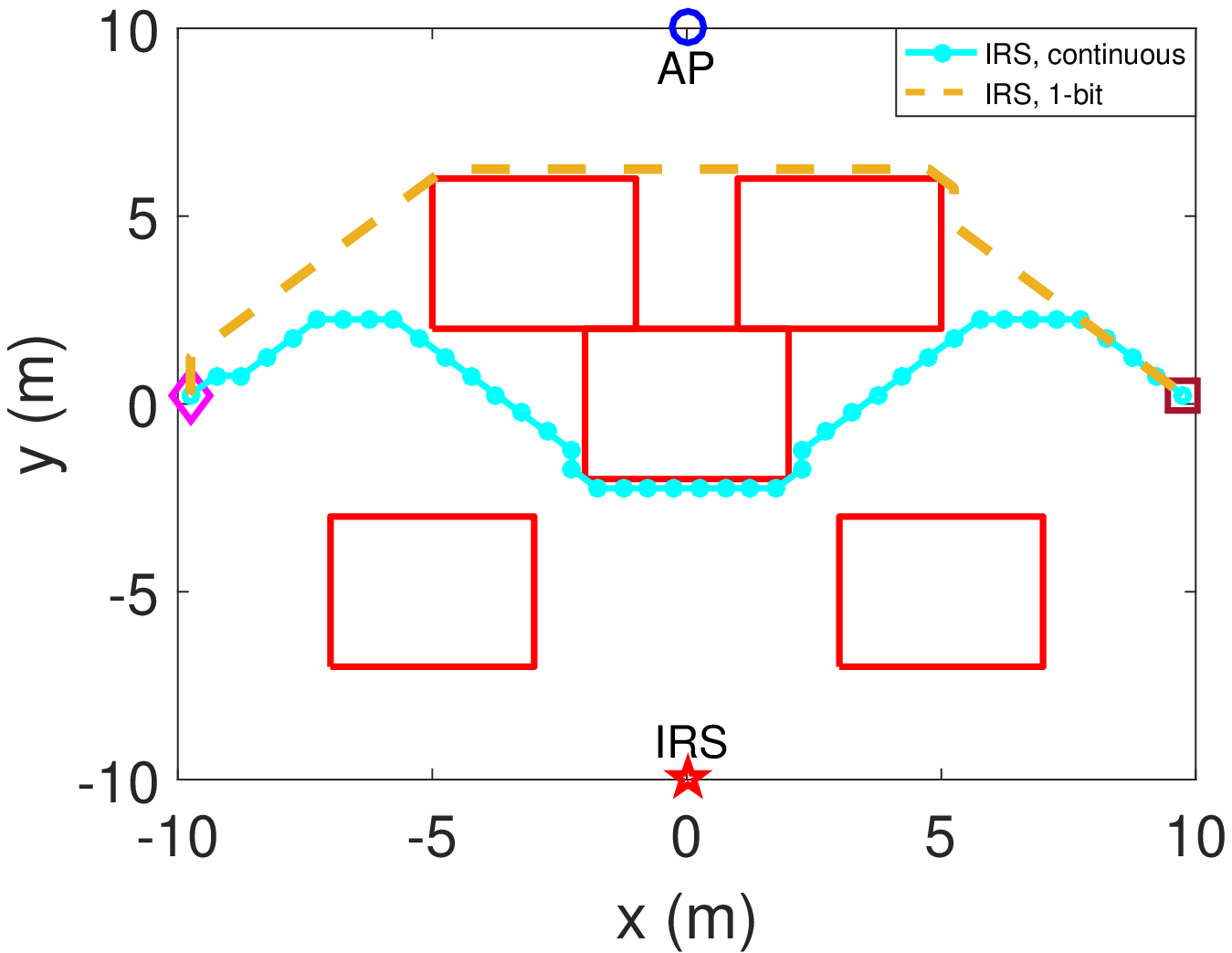}}
\setlength{\abovecaptionskip}{-0cm}
\caption{The obtained path solutions of the MRU for different $\overline \gamma$ and $M=1200$.}\label{RU path1}
\end{figure*}

Fig. \ref{RU path1} depicts the obtained paths of the MRU for different expected channel power gain targets. The red boxes represent the regions covered by obstacles. The initial and final locations of the MRU are denoted by ``$\lozenge$'' and ``$\square$'', respectively. For comparison, results without IRS and for 1-bit quantization are also shown. As can be observed in Fig. \ref{db63}, for $\overline \gamma = -63$ dB, the path obtained for the case without IRS approaches the AP to avoid the blockage caused by the obstacles. This is expected since only travelling along such a path can create a good channel condition for the MRU, which in turn leads to a longer travelling distance. However, for the IRS-aided schemes, the MRU tends to travel in a relatively straight line from ${\mathbf{q}}_I$ to ${\mathbf{q}}_F$, which leads to a shorter travelling distance compared to the case without IRS. Though the communication link between the MRU and the AP may be blocked by obstacles, a reflected LoS dominated communication link can be established with the IRS. Therefore, the MRU is not forced to travel towards the AP, since the IRS offers more degrees of freedom for path planning. This clearly demonstrates the benefits of deploying an IRS.\\
\indent In Fig. \ref{db62}, we increase $\overline \gamma$ to $-62.1$ dB. In this case, the path obtained for 1-bit quantization becomes identical to that without IRS. However, the path obtained with the IRS with continuous phase shifts still remains the same as in Fig. \ref{db63}. This is because the performance degradation caused by discrete phase shifts causes some cells to become infeasible even if they are covered by the IRS. As a result, the path planning has to mainly rely on the AP. In Fig. \ref{db61}, the expected channel power gain target is increased further to $\overline \gamma = -61$ dB. In this case, the path planning problem becomes infeasible without IRS. For 1-bit quantization, the MRU still needs to travel the longer distance around the AP to meet the expected channel power gain requirement. For the IRS with continuous phase shifts, the MRU tends to travel to regions, which are covered by the IRS through a LoS dominated communication link.
\subsubsection{Travelling Distance versus Expected Channel Power Gain Target} 
\begin{figure}[t!]
\centering
\begin{minipage}[t]{0.45\linewidth}
\includegraphics[width=2.8in]{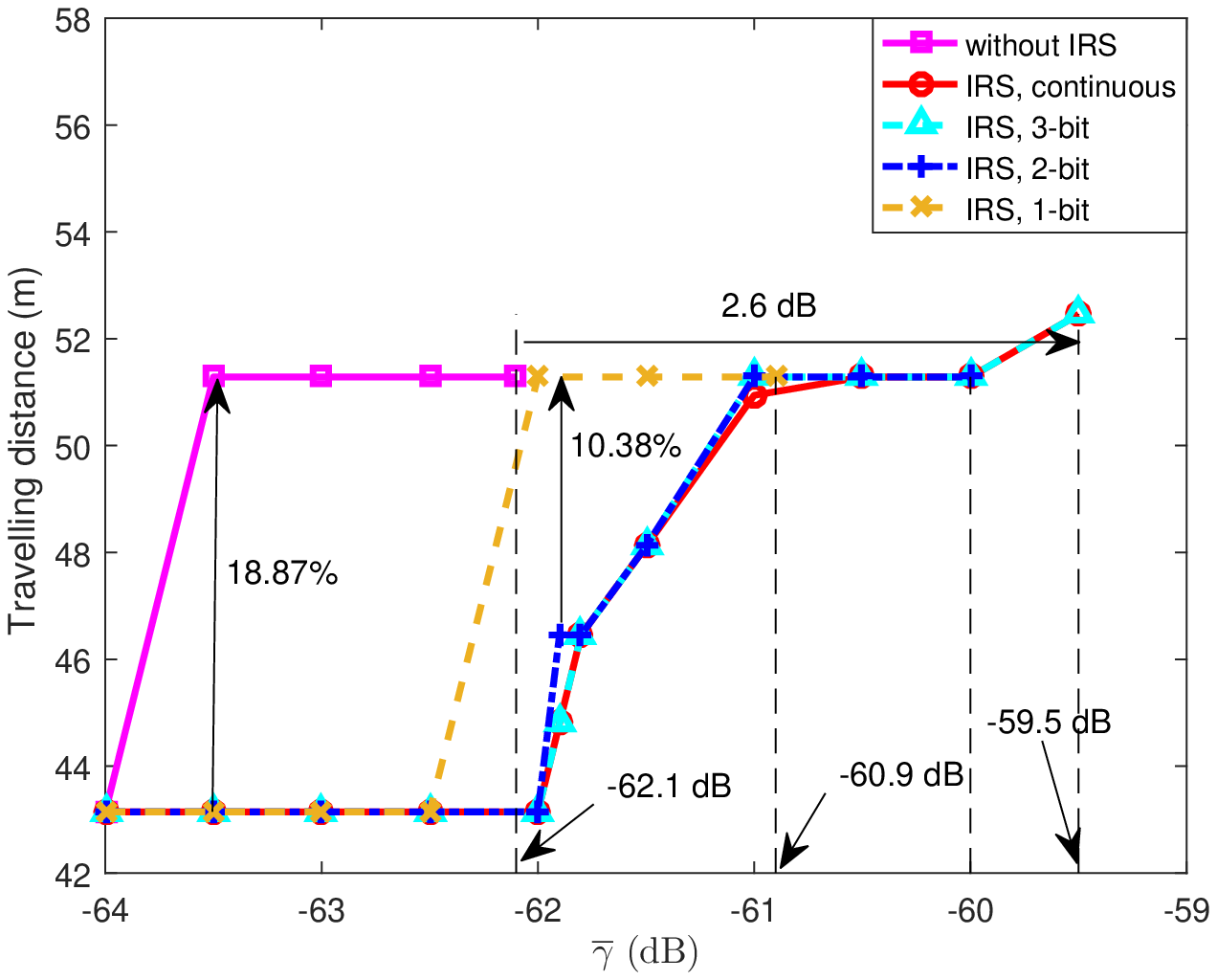}
\caption{Travelling distance versus $\overline \gamma$ for $M=1200$.}
\label{DvT}
\end{minipage}
\quad
\begin{minipage}[t]{0.45\linewidth}
\includegraphics[width=2.8in]{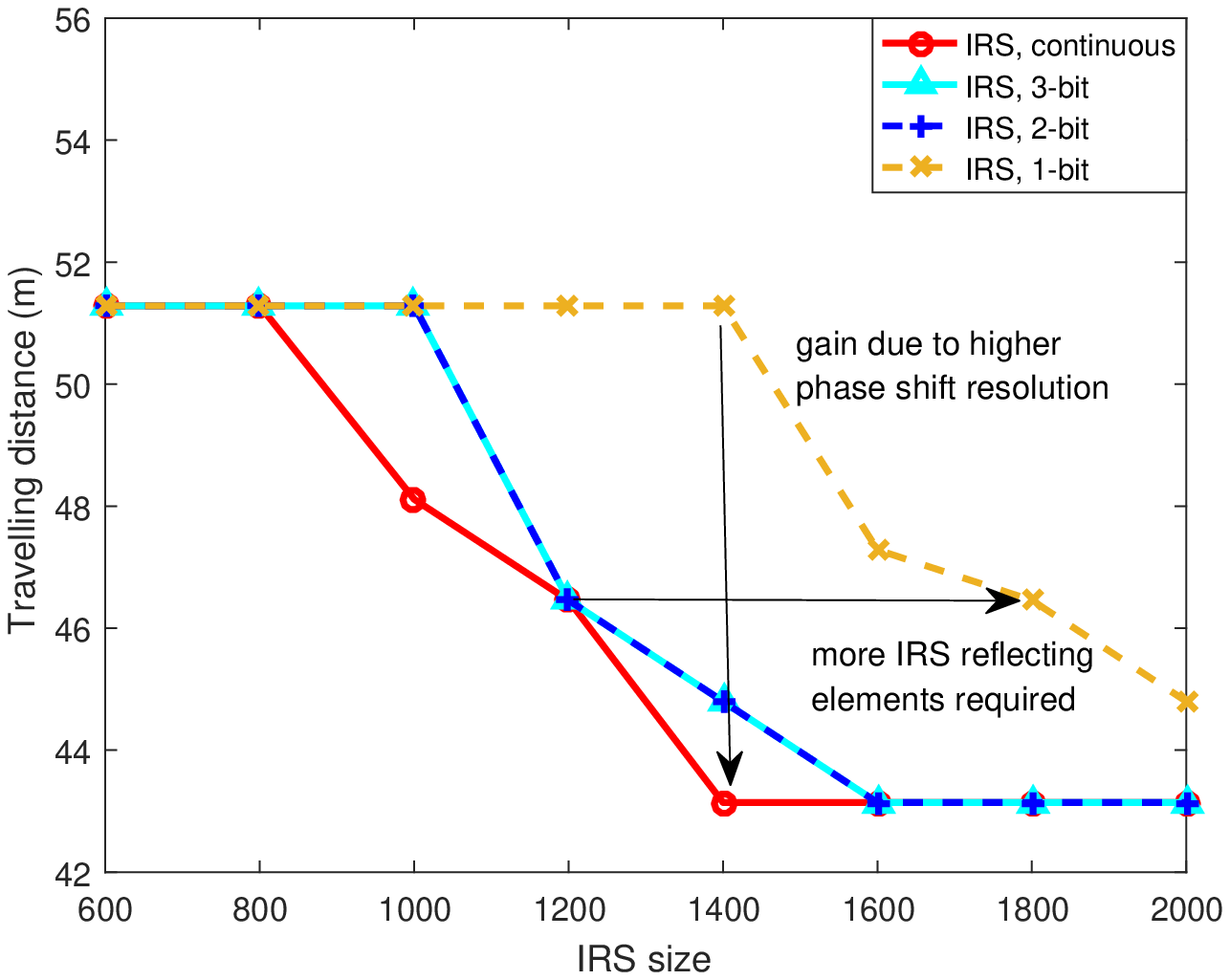}
\caption{Travelling distance versus $M$ for $\overline \gamma=-61.8$ dB.}
\label{DvS}
\end{minipage}
\end{figure}

In Fig. \ref{DvT}, we depict the travelling distance of different schemes versus the expected channel power gain target $\overline \gamma$ for $M=1200$. It is first observed that the minimum required travelling distances of all schemes generally increase as $\overline \gamma$ increases. This is expected since a larger expected channel power gain requirement reduces the number of feasible cells in the radio map, which also reduces the flexibility in path planning. Note that without IRS, the path planning problem becomes infeasible for $\overline \gamma \ge -62.1$ dB. The feasibility threshold for the IRS-aided schemes increases to $-62.1$ dB, $-60.9$ dB, $-60$ dB, and $-59.5$ dB as the phase shift resolution improves. The proposed scheme with continuous phase shifts yields a 2.6 dB performance gain over the scheme without IRS. Moreover, without IRS, when  $- 63.5\; {\rm{dB}}\le \overline \gamma   \le - 62.5$ dB the MRU needs to travel up to 18.87\% farther than when the IRS is present. Furthermore, with the 1-bit phase shifter, the required travelling distance is at most 10.38\% larger than for higher phase shift resolutions when $- 61.9\; {\rm{dB}}\le \overline \gamma   \le - 60.9$ dB. The performance degradation caused by 2- or 3-bit phase shifters is negligible compared to continuous phase shifters, which is also consistent with the results in Fig. \ref{FR}.
\subsubsection{Travelling Distance versus Number of IRS Elements} In Fig. \ref{DvS}, the required travelling distance for different schemes versus the number of IRS elements $M$ is presented. We set the expected channel power gain target to $\overline \gamma=-61.8$ dB. As can be observed, in general, the minimum required travelling distance of each scheme decreases as $M$ increases. This is because a larger number of IRS elements is capable of achieving a higher array gain, which allows the MRU to travel in a more flexible manner. Furthermore, to achieve the same travelling distance, the 1-bit phase shifter requires at most 600 additional IRS elements compared to the other phase shifters. The performance achieved by 2- or 3-bit phase shifters is close to that for continuous phase shifters. This reveals an interesting trade-off between the number of IRS elements and the number of phase shift resolution bits. Though a smaller number of phase shift resolution bits reduces the cost of the IRS elements, it increases the required number of IRS elements to achieve a certain performance, which in turn increases the deployment cost.
\vspace{-0.6cm}
\subsection{Multiple-user Scenario}
\vspace{-0.1cm}
In this subsection, we consider the multiple-user scenario. Considering the setup in Fig. \ref{setup}, the SRU's antenna is located at $\left( { 0,0,1.3} \right)$ meters, such that a LoS dominated communication link to the AP and the IRS always exists. The minimum required communication rate of the SRU is set as ${\overline r _s} = 1$ bit/s/Hz. For comparison, we consider the following benchmark scheme:
\begin{itemize}
  \item \textbf{``Z'' without IRS}: In this case, the AP serves multiple users without the help of an IRS. The communication rate map is obtained by solving problems \eqref{rate map NOMA} or \eqref{rate map OMA} while only considering the AP-user channels. Here, ``Z'' refers to NOMA or OMA.
\end{itemize}
\begin{figure*}[t!]
\centering
\subfigure[OMA without IRS]{\label{ratemapWithout}
\includegraphics[width= 1.5in, height=1.2in]{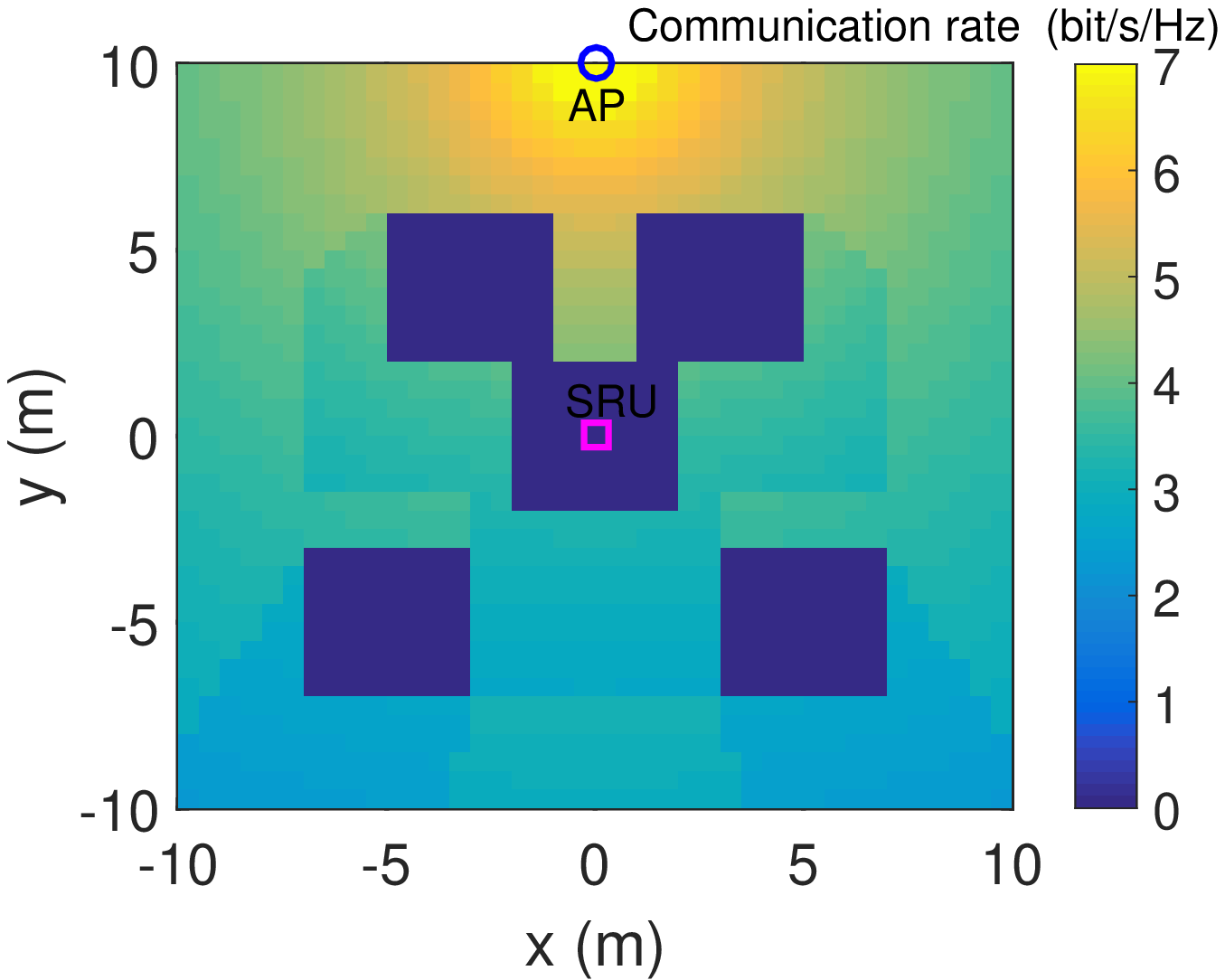}}
\subfigure[NOMA without IRS]{\label{ratemapWithoutN}
\includegraphics[width= 1.5in, height=1.2in]{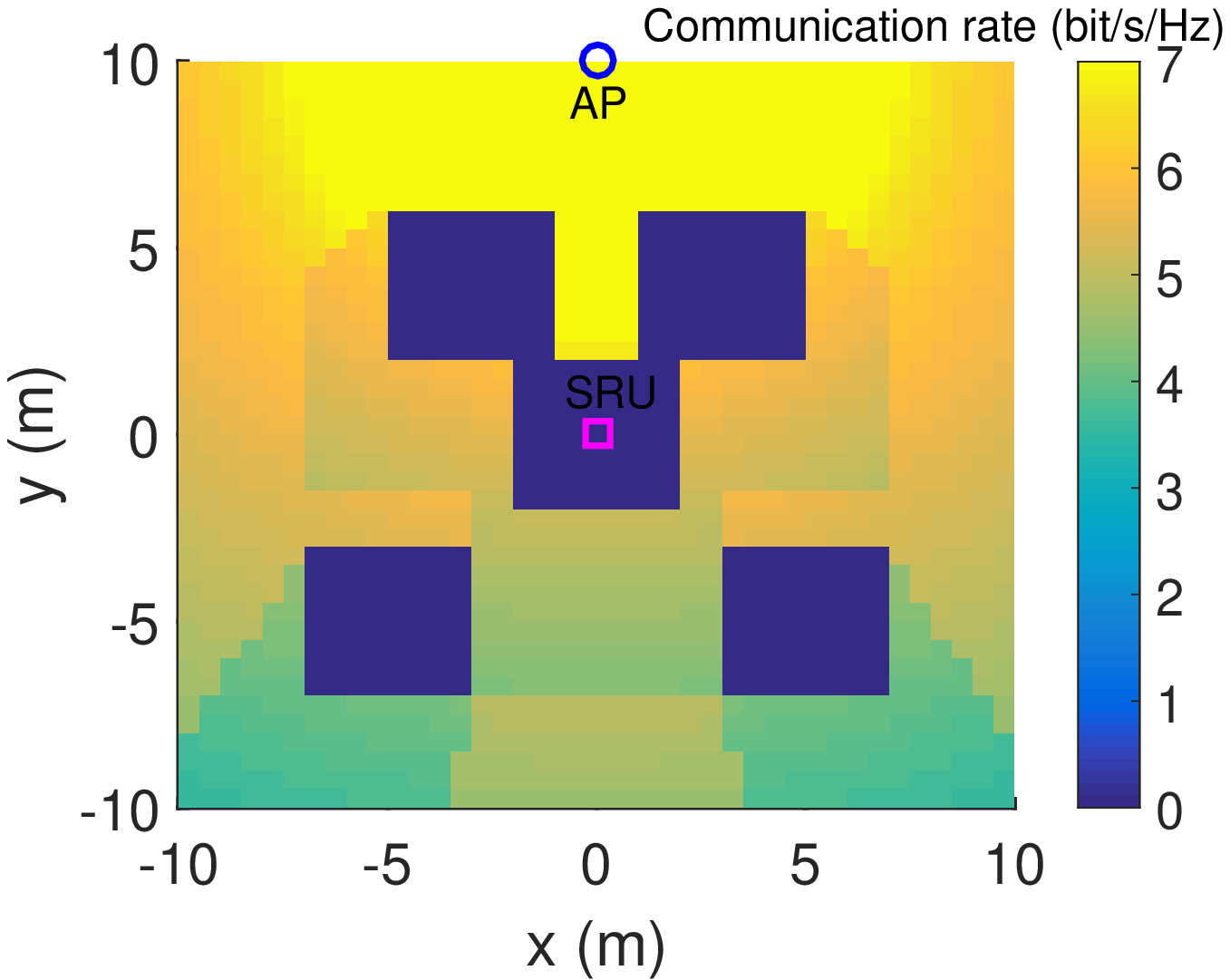}}
\subfigure[OMA with IRS]{\label{ratemapOMA1}
\includegraphics[width= 1.5in, height=1.2in]{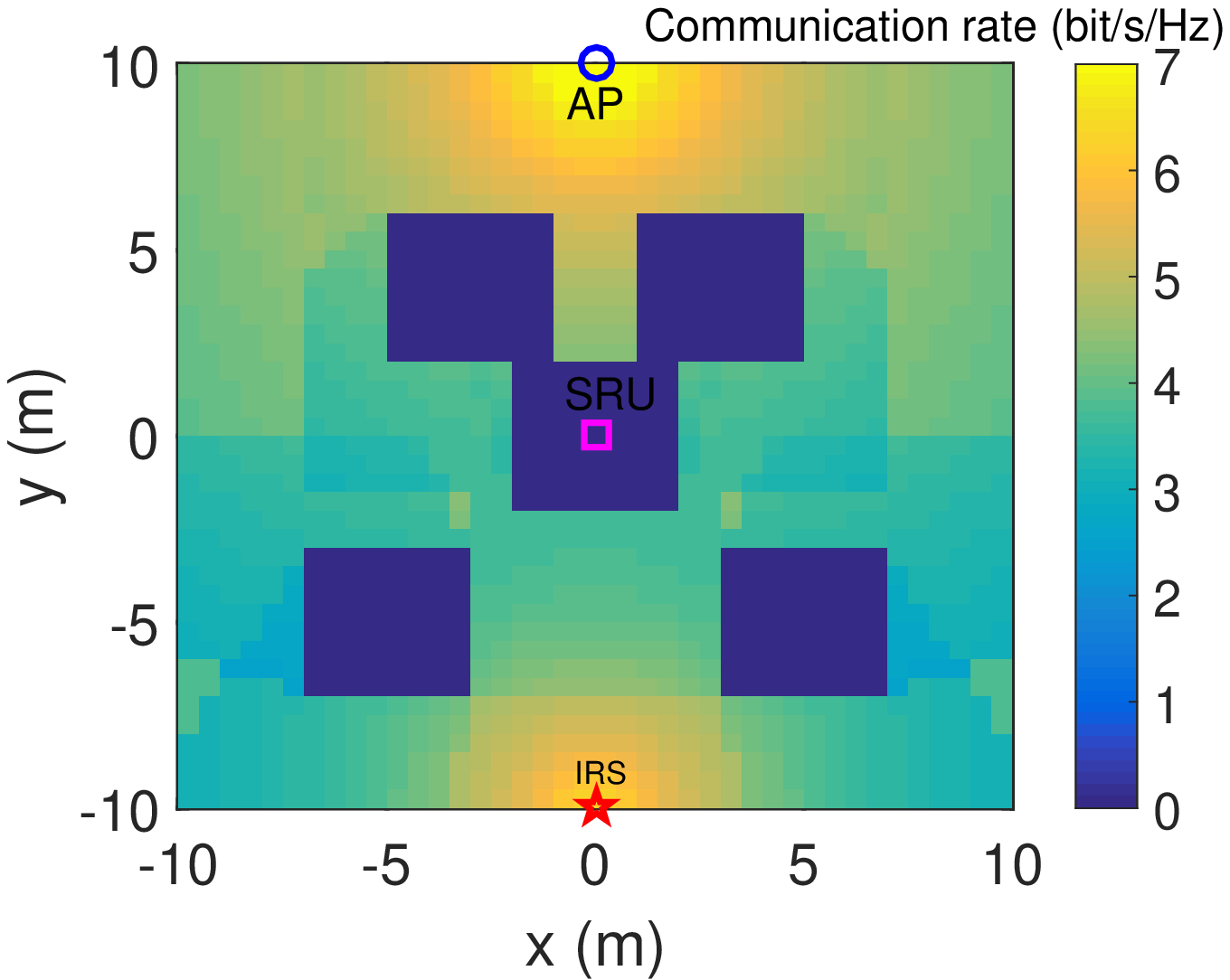}}
\subfigure[NOMA with IRS]{\label{ratemapNOMA1}
\includegraphics[width= 1.5in, height=1.2in]{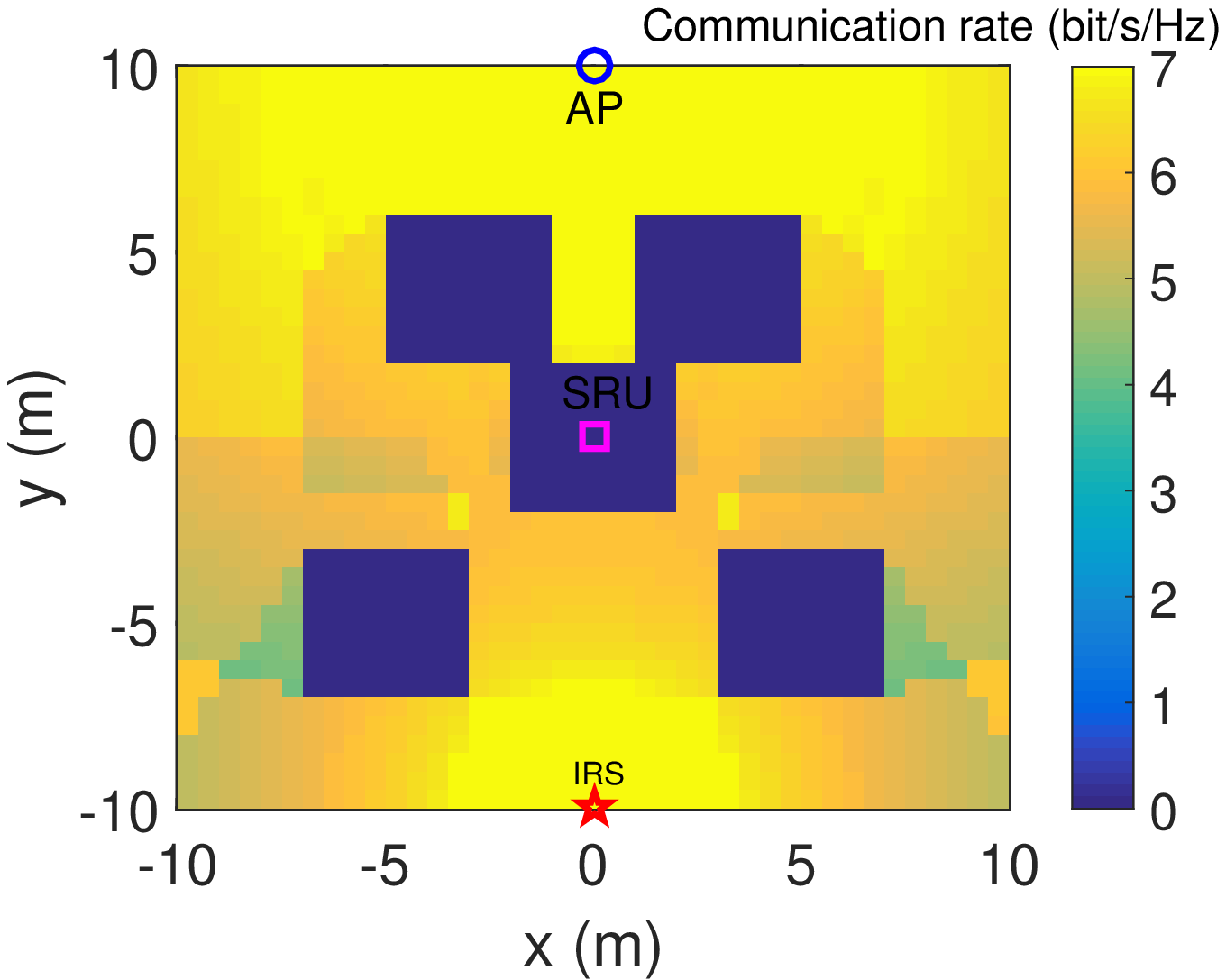}}
\setlength{\abovecaptionskip}{-0cm}
\caption{Illustration of the obtained communication rate map for different schemes and \textcolor{black}{${\Delta_x }={\Delta_y }=0.5$ m}.}\label{rate map}
\end{figure*}
\subsubsection{Communication Rate Map} Fig. \ref{rate map} depicts the obtained communication rate map for different schemes and $M=1200$. The location of the SRU is denoted by ``$\square$''. As shown in Fig. \ref{ratemapWithout}, with OMA, only a small region can achieve a rate of more than 5 bit/s/Hz for the MRU, if an IRS is not present. However, in Fig. \ref{ratemapWithoutN}, it can be observed that more than half of the cells can achieve a rate of more than 5 bit/s/Hz if NOMA is employed even without the help of an IRS. This is because NOMA allows the two users to share their resource blocks, which improves spectrum efficiency. A significant rate degradation can still be observed in the regions behind the obstacles for both multiple access schemes due to blockage. Fig. \ref{ratemapOMA1} and Fig. \ref{ratemapNOMA1} show that deploying an IRS significantly improves the expected communication rate, especially for the cells around the IRS. The rate improvement introduced by the IRS is more pronounced for NOMA compared to OMA. In fact, with NOMA, the MRU can achieve a rate of 5 bit/s/Hz or more in 90\% of the cells. The rate loss caused by blockages is reduced in the IRS-aided systems since an additional reflected LoS dominated communication link can be established via the IRS.
\subsubsection{Obtained Path of the MRU} 
\begin{figure*}[t!]
\centering
\subfigure[$\overline r_m = 3.3$ bit/s/Hz]{\label{p33}
\includegraphics[width= 2in, height=1.6in]{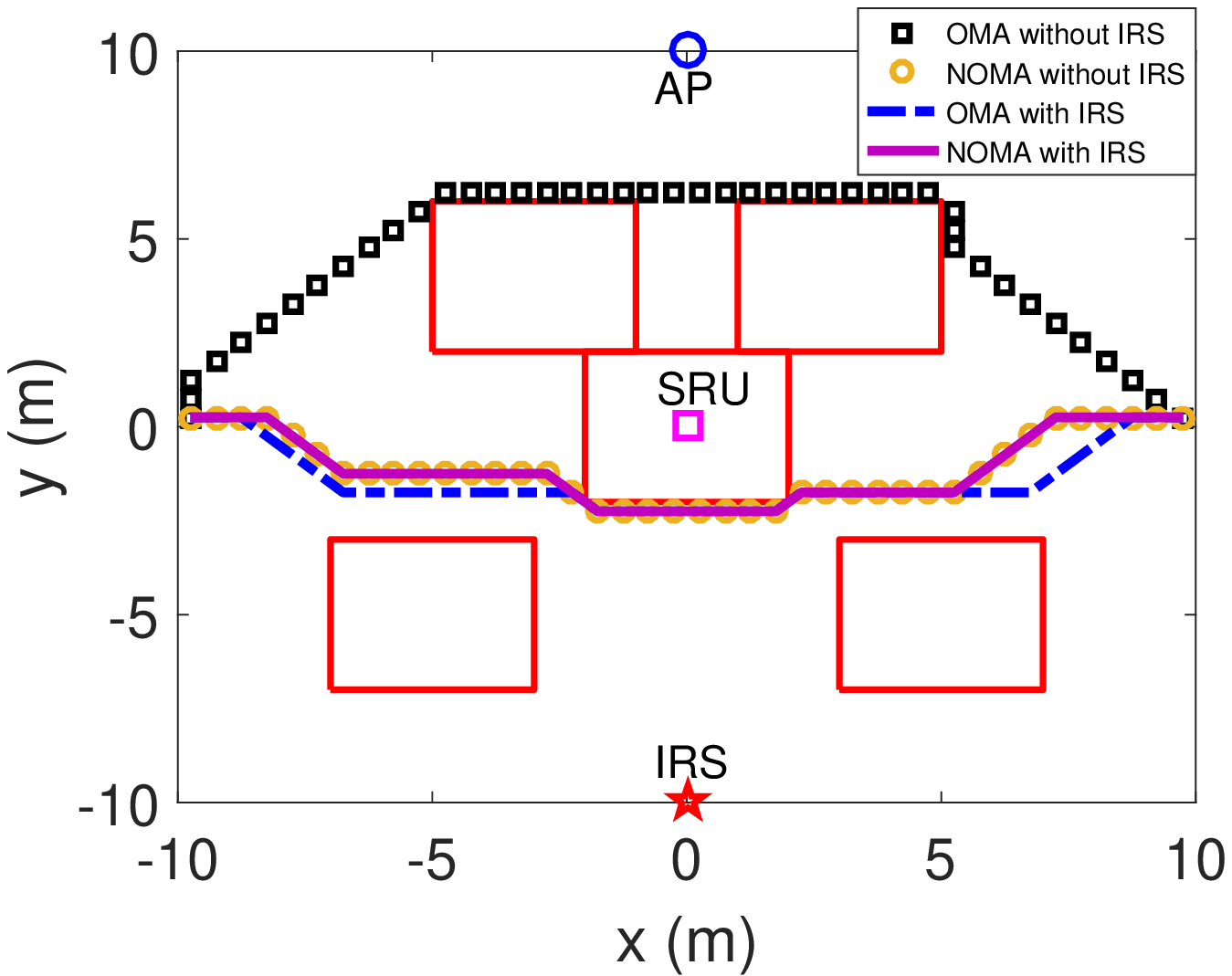}}
\subfigure[$\overline r_m = 3.6$ bit/s/Hz]{\label{p36}
\includegraphics[width= 2in, height=1.6in]{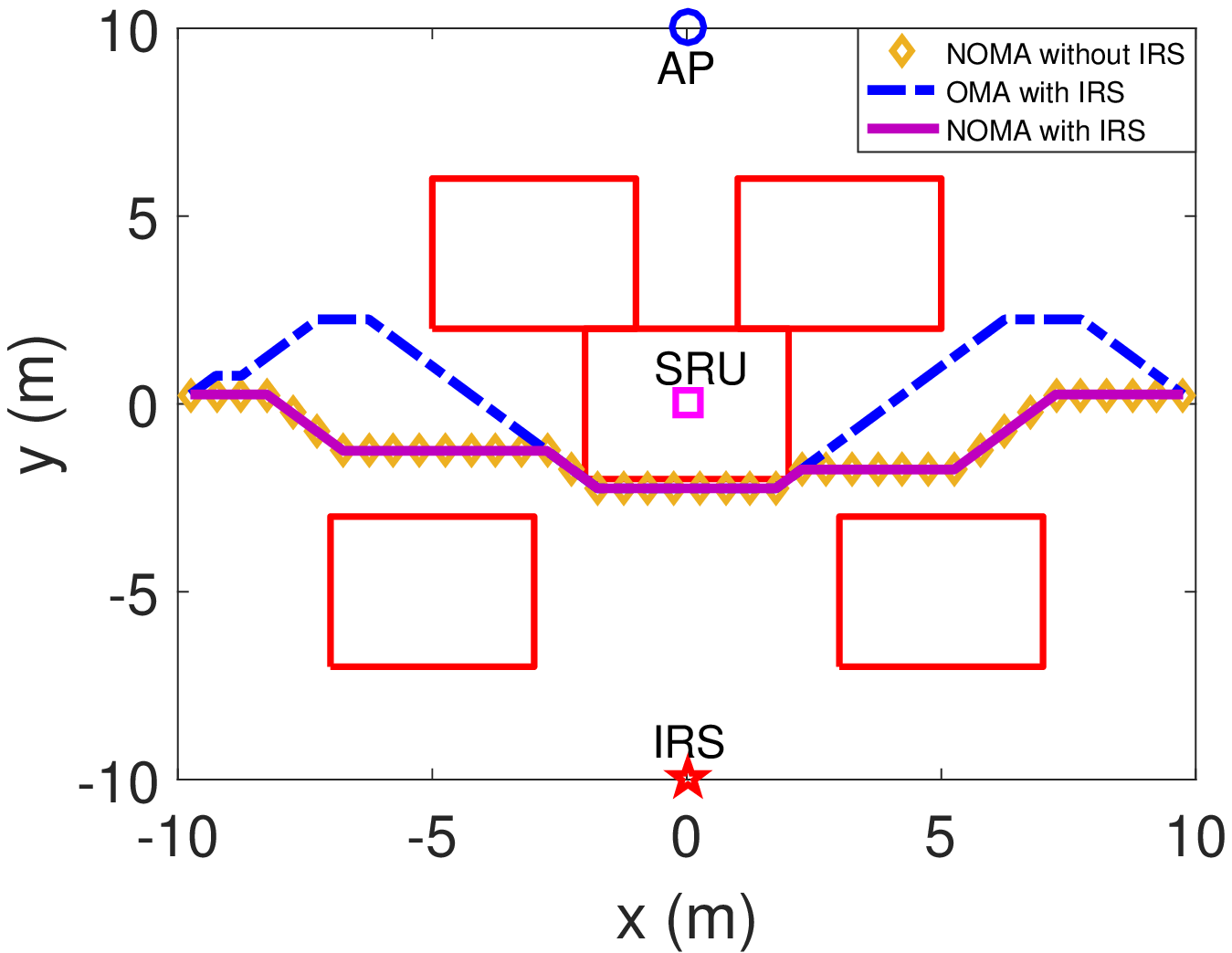}}
\subfigure[$\overline r_m = 5.3$ bit/s/Hz]{\label{p53}
\includegraphics[width= 2in, height=1.6in]{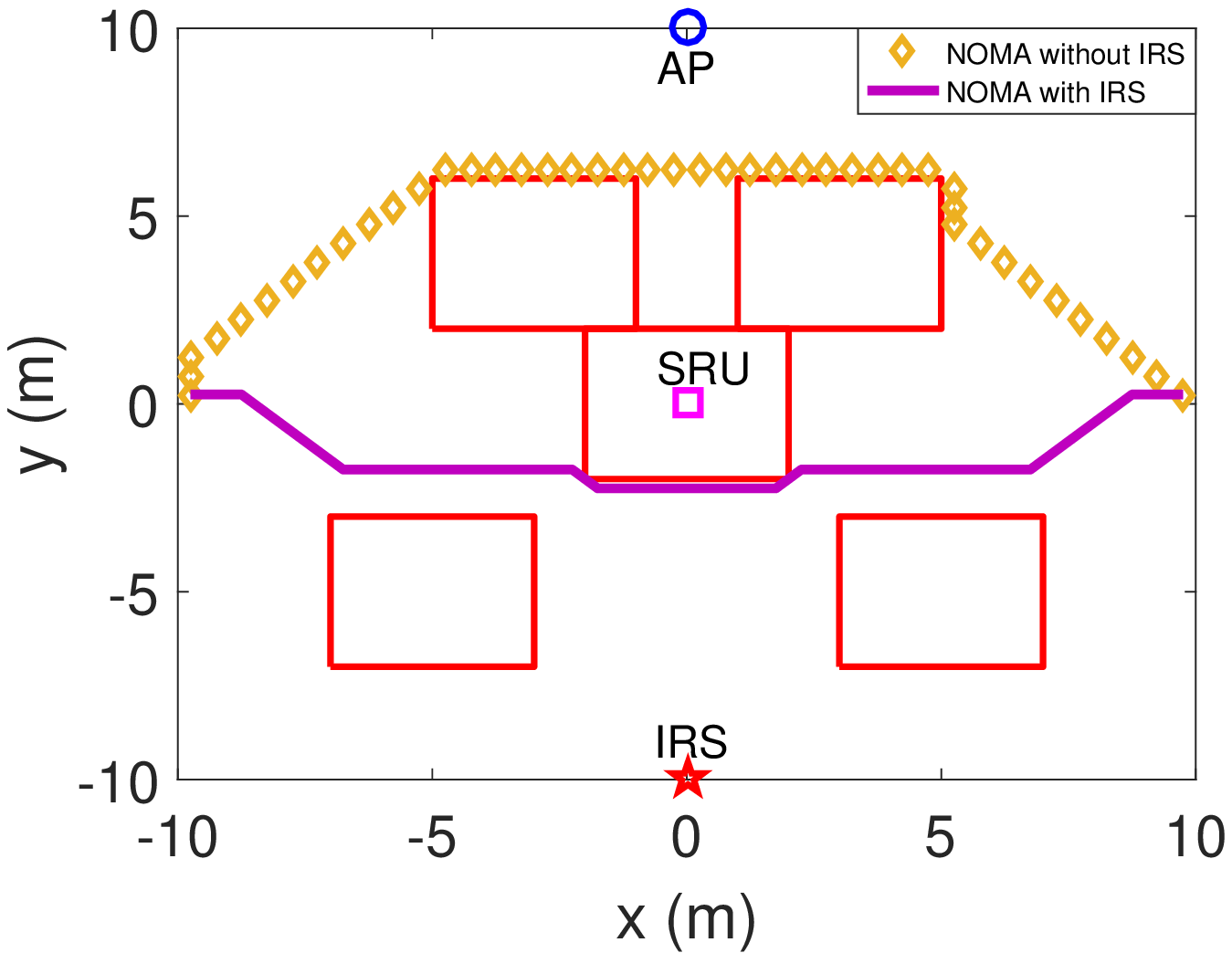}}
\setlength{\abovecaptionskip}{-0cm}
\caption{Paths of MRU for different schemes for given $\overline r_m$ and $M=1200$.}\label{RP2}
\end{figure*}

Based on the constructed communication rate map, we plot in Fig. \ref{RP2} the obtained paths for the MRU for different schemes. As shown in Fig. \ref{p33}, when $\overline r_m=3.3$ bit/s/Hz, the MRU needs to take the longer path around the AP only for OMA without IRS, while it takes a more direct path from ${\mathbf{q}}_I$ to ${\mathbf{q}}_F$ for the other three schemes. This demonstrates the benefits of NOMA and deploying an IRS. In Fig. \ref{p36}, we increase the required rate from $\overline r_m=3.3$ bit/s/Hz to $\overline r_m=3.6$ bit/s/Hz. In this case, the path planning problem becomes infeasible for OMA without IRS. For NOMA with and without IRS, the path of the MRU remains unchanged, while the path for OMA with IRS tends to traverse the cells covered by the IRS exploiting the reflected LoS dominated channel. In Fig. \ref{p53}, where $\overline r_m=5.3$ bit/s/Hz, the path planning problem becomes infeasible if OMA is used. For NOMA without IRS, the MRU has to approach the AP to achieve the required communication rate, which increases the travelling distance. For NOMA with IRS, the path remains unchanged compared to Fig. \ref{p36}. This underscores the effectiveness of the proposed IRS-aided NOMA scheme.
\subsubsection{Travelling Distance versus Required Communication Rate Target} In Fig. \ref{DvR}, we show the travelling distance versus the required rate target $\overline r_m$ for $M=1200$. We first observe that the travelling distance increases as the required rate target $\overline r_m$ increases. Without IRS, the path planning problem becomes infeasible when $\overline r_m \ge 3.45$ bit/s/Hz for OMA and $\overline r_m \ge 5.3$ bit/s/Hz for NOMA. With IRS, the threshold increases to 3.9 bit/s/Hz for OMA and 6.25 bit/s/Hz for NOMA. With IRS, the gain of NOMA over OMA is more pronounced than without IRS. This is because properly configured the IRS phase shifts can enhance the channel disparity between the two users, which benefits NOMA. Furthermore, it is also observed that for NOMA the IRS gain is more pronounced than for OMA. This implies that deploying an IRS is more beneficial if NOMA is employed.
\subsubsection{Tightness of Expected Achievable Rate Approximation} 
\begin{figure}[t!]
\centering
\begin{minipage}[t]{0.45\linewidth}
\includegraphics[width=2.8in]{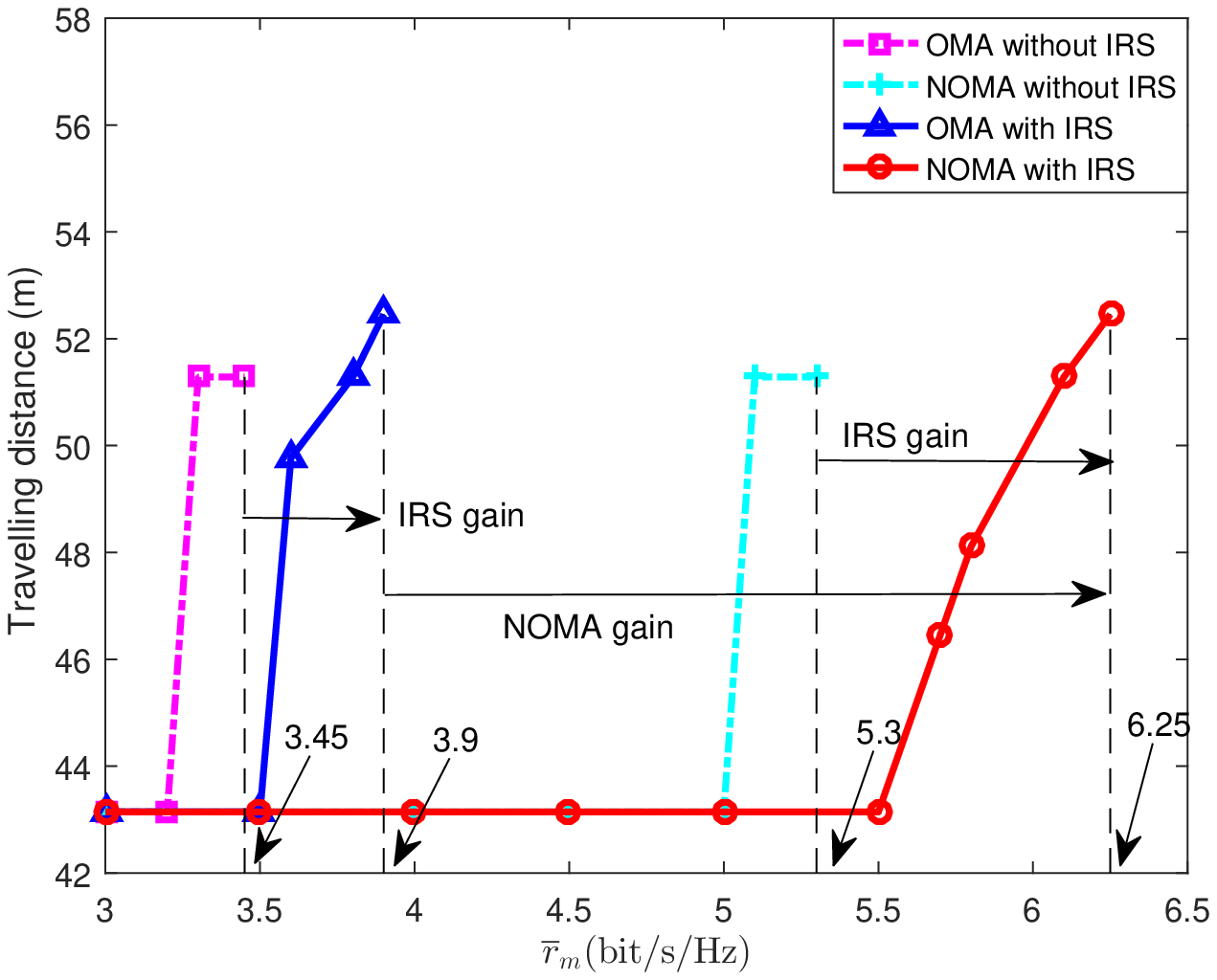}
\caption{Travelling distance versus $\overline r_m$ for $M=1200$.}
\label{DvR}
\end{minipage}
\quad
\begin{minipage}[t]{0.45\linewidth}
\includegraphics[width=2.8in]{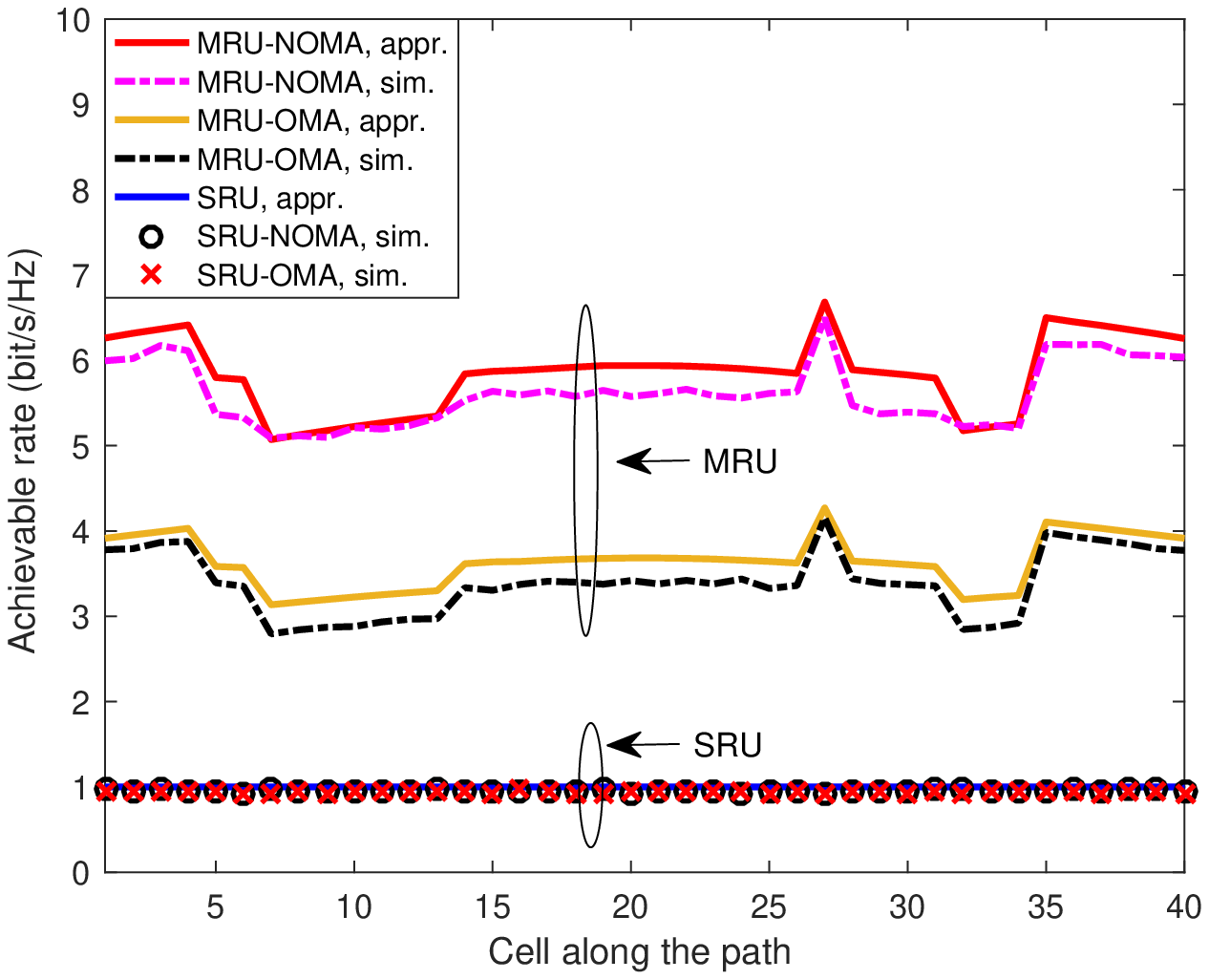}
\caption{Comparison of approximate and the exact expected achievable rate.}
\label{app}
\end{minipage}
\end{figure}

In Fig. \ref{app}, we evaluate the tightness of the approximation of the expected achievable rate. Specifically, the MRU is assumed to travel along the path in Fig. \ref{p33} for OMA and NOMA. For each cell along the path, the approximation of the expected achievable rates, i.e., ${\overline R _l^{\rm{NOMA}}}\left( {{\mathbf{q}}_{i,j}^\Delta } \right)$ and ${\overline R _l^{\rm{OMA}}}\left( {{\mathbf{q}}_{i,j}^\Delta } \right)$, are calculated with \eqref{approximated rate NOMA} and \eqref{approximated rate OMA}. The exact expected achievable rates, i.e., ${\mathbb{E}}\left[ {R_l^{\rm{NOMA}}\left( {{\mathbf{q}}_{i,j}^\Delta } \right)} \right]$ and ${\mathbb{E}}\left[ {R_l^{\rm{OMA}}\left( {{\mathbf{q}}_{i,j}^\Delta } \right)} \right]$, are obtained via Monte Carlo simulation by averaging over 10000 random channel realizations for each cell. As can be observed, the approximations match well with the exact results for the SRU for OMA and NOMA. For the MRU, a small gap can be observed between the approximation and the exact results since the approximation is an upper bound for the exact average achievable rate. This implies that, in a practical implementation, a small constant should be added to the required rate ${\overline r_m}$ in the proposed optimization problem \eqref{PNOMA} to account for the gap between the upper bound and the actual average achievable rate.
\vspace{-0.3cm}
\section{Conclusions}
\vspace{-0.1cm}
An IRS-assisted indoor robot navigation system has been investigated. The communication-aware robot path planning problem was formulated for minimization of the travelling time/distance by jointly optimizing the robot path and the phase shifts of the IRS elements. To solve this problem, we proposed a radio map based approach which exploits knowledge about the location-dependent channel propagation. Channel power gain maps and the communication rate maps were constructed for single-user and multiple-user systems, respectively. Based on these two radio maps, the robot path planning problem was efficiently solved by invoking graph theory. Numerical results showed that the coverage of the AP can be significantly extended by deploying an IRS, and the robot travelling distance can be significantly reduced with the aid of an IRS and NOMA.\\
\indent This paper assumed perfect knowledge of the geographic information of the considered indoor environments, which can be difficult to obtain in some applications (e.g., search and rescue missions). An important direction for future research is to investigate communication-aware robot path planning in uncertain environments. In this case, simultaneous localization and mapping (SLAM) \cite{SLAM} may be a promising approach to assist radio map construction.
\vspace{-0.2cm}
\section*{Appendix~A: Proof of Lemma~\ref{expected effective channel power gain}} \label{Appendix:A}
\vspace{-0.2cm}
The expected effective channel power gain of the MRU, ${\mathbb{E}}\left[ {{{\left| {{c_m}\left( t \right)} \right|}^2}} \right]$, can be decomposed as follows:
\vspace{-0.5cm}
\begin{align}\label{x0}
\begin{gathered}
  {\mathbb{E}}\left[ {{{\left| {{c_m}\left( t \right)} \right|}^2}} \right] = {\mathbb{E}}\left\{ {{{\left| {\left( {\widetilde h_m^H\left( {{\mathbf{q}}\left( t \right)} \right) + \breve h_m^H} \right) + \left( {\widetilde {\mathbf{r}}_m^H\left( {{\mathbf{q}}\left( t \right)} \right) + \breve {\mathbf{r}}_m^H} \right){\mathbf{\Theta}} \left( t \right)\left( {\widetilde {\mathbf{g}} + \breve {\mathbf{g}}} \right)} \right|}^2}} \right\}  \hfill \\
  \mathop  = \limits^{\left( a \right)} {\left| {{x_1}} \right|^2} + {\mathbb{E}}\left\{ {{{\left| {{x_2}} \right|}^2}} \right\} + {\mathbb{E}}\left\{ {{{\left| {{x_3}} \right|}^2}} \right\} + {\mathbb{E}}\left\{ {{{\left| {{x_4}} \right|}^2}} \right\} + {\mathbb{E}}\left\{ {{{\left| {{x_5}} \right|}^2}} \right\}, \hfill \\
\end{gathered}
\end{align}
\vspace{-0.4cm}

\noindent where $\widetilde h_m^H\left( {{\mathbf{q}}\left( t \right)} \right) = \sqrt {\frac{{{{{\mathcal{L}}_{AM}}\left( {{\mathbf{q}}\left( t \right)} \right)}{K_{AM}}\left( {{\mathbf{q}}\left( t \right)} \right)}}{{{K_{AM}}\left( {{\mathbf{q}}\left( t \right)} \right) + 1}}} \overline h_m^H{\left( {{\mathbf{q}}\left( t \right)} \right)}$, $\breve h_m^H = \sqrt {\frac{{{{{\mathcal{L}}}_{AM}}\left( {{\mathbf{q}}\left( t \right)} \right)}}{{{K_{AM}}\left( {{\mathbf{q}}\left( t \right)} \right) + 1}}} \widehat h_m^H$, ${\widetilde {\mathbf{r}}_m^H}\left( {{\mathbf{q}}\left( t \right)} \right) = $\\$\sqrt {\frac{{{{\mathcal{L}}}_{IM}}\left( {{\mathbf{q}}\left( t \right)} \right){{K_{IM}}\left( {{\mathbf{q}}\left( t \right)} \right)}}{{{K_{IM}}\left( {{\mathbf{q}}\left( t \right)} \right) + 1}}} {\overline {\mathbf{r}} _m^H}\left( {{\mathbf{q}}\left( t \right)} \right)$, $\breve {\mathbf{r}}_m^H =\sqrt {\frac{{{{{\mathcal{L}}}_{IM}}\left( {{\mathbf{q}}\left( t \right)} \right)}}{{{K_{IM}}\left( {{\mathbf{q}}\left( t \right)} \right) + 1}}} {\widehat {\mathbf{r}}_m^H} $, $\widetilde {\mathbf{g}} = \sqrt {\frac{{{{{\mathcal{L}}}_{AI}}{K_{AI}}}}{{{K_{AI}} + 1}}} \overline {\mathbf{g}} $, and $\breve {\mathbf{g}} = \sqrt {\frac{{{{{\mathcal{L}}}_{AI}}}}{{{K_{AI}} + 1}}} \widehat {\mathbf{g}}$. In \eqref{x0}, $\left( a \right)$ is due to the fact that $\breve h_m^H$, $\breve {\mathbf{r}}_m^H$, and $\breve {\mathbf{g}}$ have zero means and are independent from each other. We have
\vspace{-0.6cm}
\begin{subequations}
\begin{align}
\label{x1}&{\left| {{x_1}} \right|^2} = {\left| {\widetilde h_m^H\left( {{\mathbf{q}}\left( t \right)} \right) + \widetilde {\mathbf{r}}_m^H\left( {{\mathbf{q}}\left( t \right)} \right){\mathbf{\Theta}} \left( t \right)\widetilde {\mathbf{g}}} \right|^2},\\
\label{x2}&{\mathbb{E}}\left\{ {{{\left| {{x_2}} \right|}^2}} \right\} = {\mathbb{E}}\left\{ {{{\left| \breve {h_m^H} \right|}^2}} \right\} = \frac{{{\mathcal{L}}_{AM}}\left( {{\mathbf{q}}\left( t \right)} \right)}{{{K_{AM}}\left( {{\mathbf{q}}\left( t \right)} \right) + 1}},\\
\label{x3}&{\mathbb{E}}\left\{ {{{\left| {{x_3}} \right|}^2}} \right\} = {\mathbb{E}}\left\{ {{{\left| {\widetilde {\mathbf{r}}_m^H\left( {{\mathbf{q}}\left( t \right)} \right){\mathbf{\Theta}} \left( t \right)\breve {\mathbf{g}}} \right|}^2}} \right\} = \frac{{{{{\mathcal{L}}}_{AI}}{{{\mathcal{L}}}_{IM}}\left( {{\mathbf{q}}\left( t \right)} \right){K_{IM}}\left( {{\mathbf{q}}\left( t \right)} \right)M}}{{\left( {{K_{AI}} + 1} \right)\left( {{K_{IM}}\left( {{\mathbf{q}}\left( t \right)} \right) + 1} \right)}},\\
\label{x4}&{\mathbb{E}}\left\{ {{{\left| {{x_4}} \right|}^2}} \right\} = {\mathbb{E}}\left\{ {{{\left| {\breve {\mathbf{r}}_m^H\left( {{\mathbf{q}}\left( t \right)} \right){\mathbf{\Theta}} \left( t \right)\widetilde {\mathbf{g}}} \right|}^2}} \right\} = \frac{{{{{\mathcal{L}}}_{AI}}{K_{AI}}{{{\mathcal{L}}}_{IM}}\left( {{\mathbf{q}}\left( t \right)} \right)M}}{{\left( {{K_{AI}} + 1} \right)\left( {{K_{IM}}\left( {{\mathbf{q}}\left( t \right)} \right) + 1} \right)}},\\
\label{x5}&{\mathbb{E}}\left\{ {{{\left| {{x_5}} \right|}^2}} \right\} = {\mathbb{E}}\left\{ {{{\left| {\breve {\mathbf{r}}_m^H\left( {{\mathbf{q}}\left( t \right)} \right){\mathbf{\Theta}} \left( t \right)\breve {\mathbf{g}}} \right|}^2}} \right\}  = \frac{{{{{\mathcal{L}}}_{AI}}{{{\mathcal{L}}}_{IM}}\left( {{\mathbf{q}}\left( t \right)} \right)M}}{{\left( {{K_{AI}} + 1} \right)\left( {{K_{IM}}\left( {{\mathbf{q}}\left( t \right)} \right) + 1} \right)}}.
\end{align}
\end{subequations}
\vspace{-1cm}

\indent Therefore, by inserting the results in \eqref{x1}-\eqref{x5} into \eqref{x0}, we arrive at \eqref{expected robotic channel gain0}. This completes the proof of Lemma 1.

\vspace{-0.6cm}
\bibliographystyle{IEEEtran}
\bibliography{mybib}

\end{document}